\documentclass[twocolumn,showpacs,preprintnumbers,amsmath,amssymb,superscriptaddress]{revtex4}

\usepackage{graphicx}
\usepackage{dcolumn}
\usepackage{bm}
\usepackage{array}

\makeatletter
\newcount\@cla
\newcount\@clb
\makeatother
\usepackage{sha}

\def\ket{\rangle}

\begin{document}

\title{Complex Energy Spectrum and Time Evolution of QBIC States\\
in a Two-Channel Quantum wire with an Adatom Impurity}

\author{S. Garmon}
 \email{stergar@physics.utexas.edu}
 \affiliation{
 Laboratoire de Physique Th\'eorique de la Mati\`ere Condens\'ee, 
 CNRS UMR 7600, 
 Universit\'e Pierre et Marie Curie,
 4 Place Jussieu, 75252 Paris Cedex 05, France}
 \affiliation{
Center for Complex Quantum Systems, University of Texas at Austin,
1 University Station, C1609, Austin, TX 78712, United States}%

\author{H. Nakamura}
 \email{hnakamura@nifs.ac.jp}
\affiliation{%
National Institute for Fusion Science,
Oroshi-cho 322-6, Toki, Gifu 509-5292, Japan}

\author{N. Hatano}%
 \email{hatano@iis.u-tokyo.ac.jp}
\affiliation{%
Institute of Industrial Science,
University of Tokyo,
Komaba 4-6-1, Meguro, Tokyo 153-8505, Japan}

\author{T. Petrosky}
 \email{petrosky@physics.utexas.edu}
\affiliation{
Center for Complex Quantum Systems, University of Texas at Austin,
1 University Station, C1609, Austin, TX 78712, United States}%

\date{\today}

\begin{abstract}
We provide detailed analysis of the complex energy eigenvalue spectrum for a two-channel quantum wire with an attached adatom impurity.  The study is based on our previous work [Phys.\ Rev.\ Lett.\ {\bf 99}, 210404 (2007)], in which we presented the quasi-bound states in continuum (or QBIC states).  These are resonant states with very long lifetimes that form as a result of two overlapping continuous energy bands one of which, at least, has a divergent van Hove singularity at the band edge.  We provide analysis of the full energy spectrum for all solutions, including the QBIC states, and obtain an expansion for the complex eigenvalue of the QBIC state.  We show that it has a small decay rate of the order $g^6$, where $g$ is the  coupling constant.  As a result of this expansion, we find that this state is a non-analytic effect resulting from the van Hove singularity; it cannot be predicted from the ordinary perturbation analysis that relies on Fermi's golden rule.  We will also numerically demonstrate the time evolution of the QBIC state using the effective potential method in order to show the stability of the QBIC wave function in comparison with that of the other eigenstates.  
\end{abstract}

\maketitle

\section{Introduction}

In a previous Letter~\cite{QBIC-Letter} we introduced the {\it quasi-bound states in continuum} (QBIC) as resonant states which occur in certain systems with overlapping continuous energy bands.  Consider the case where one of these energy bands has a divergent van Hove singularity in the density of states at one of the overlapping band edges.  Then
a discrete excited state coupled to these energy bands gives rise to a metastable resonant state with an extended lifetime.  This effect cannot be predicted using Fermi's golden rule as it breaks down in the vicinity of the singularity.

In our Letter, we demonstrated the existence of the QBIC states in the context of a two-channel quantum wire coupled to a single adatom impurity.  This built on previous work by S.~Tanaka, S.~G., and T.~P.\ on a single channel wire coupled to an adatom~\cite{06TGP}, in which they demonstrated various non-analytic effects that resulted from the presence of the divergent van Hove singularity in the electron density of states (DOS)~\cite{53vanHove}
 at the edge of the conduction band; note that these are characteristic effects of the van Hove singularity in a one-dimensional system.  One of these effects was a bound state that lies just outside of either edge of the conduction band, no matter how deeply the discrete adatom energy is embedded in the 
band~\cite{06TGP,Mahan,05PTG,SG-Diss}.  We will refer to this state here as a \emph{persistent bound state} as it co-exists with the unstable decay states.  In the present paper, we will show that when the adatom is coupled to a second energy band that overlaps the first, this persistent bound state is slightly destabilized due to the fact that it lies in the continuum of the second energy band.  We will explicitly demonstrate that it is this persistent bound state (when de-stabilized) that forms the QBIC eigenstate for the two-channel system; for this purpose, we compare term-by-term the analytic expansions of the bound state energy eigenvalue in the single channel system and that of the QBIC energy eigenvalue in the two-channel system.  We will also show that the decay rate for the QBIC state is (to the lowest order) 
proportional to  $g^6$, where $g$ is the  coupling constant between the adatom and the site to which it is attached.  We assume that $g$ is small in this paper.

We have envisioned the QBIC state as a generalization of the bound state in continuum (BIC) originally proposed by von Neumann and Wigner in 1929~\cite{29vonNeu}.  Since their initial proposal, a good deal of theoretical study has been devoted to this phenomenon~\cite{Stillinger75,Gazdy77,Fonda60,Sudarshan81,Friedrich85,OK04,Ordonez06,Sadreev06,Bulgakov06}, including a recent paper by S.~Longhi, in which the author demonstrates the presence of BIC states in a single level semi-infinite Fano-Anderson model~\cite{Longhi07} 
and another article by S.~Tanaka, S.~G., G.~Ordonez, and T.~P., in which the authors demonstrate the presence of BIC states in a single channel quantum wire coupled with multiple adatom impurities~\cite{TGOP07,SG-Diss}.  There has also been experimental confirmation of the BIC phenomenon~\cite{Capasso92,Deo94}.  However, since it is a zero measure effect (meaning that it only occurs at discrete points in parameter space), it is generally considered difficult to detect.  

As discussed in our previous Letter, the QBIC state actually has a decay rate (due to the imaginary part of the complex eigenenergy) and hence does not technically lie in the continuous energy spectrum of the conduction band.  However, the decay rate for this state is on the order of $g^6$, much smaller than the ordinary decay rate that is predicted to be of the order $g^2$ by Fermi's golden rule.  Hence the QBIC state will behave as a bound state (with real part of the complex energy inside the continuous energy spectrum) even on relatively large time scales.  We will also show below that the $g^6$ power in the decay rate is a direct result of the interaction between the divergent van Hove singularity at one of the band edges and the continuum of the other band.

While the BIC states occur only at discrete values in parameter space, the QBIC states occur over a wide range of parameter space.  Hence, it may be much easier to experimentally verify the QBIC effect.  It may also be easier to verify the QBIC effect, considering that the effect will likely occur in other physical systems; for instance, in another Letter~\cite{05PTG} by C.-O.~Ting, T.~P.\ and S.~G., the authors explored the effects of the divergent van Hove singularity in the photon density of states at the cutoff frequency 
in the interaction between an excited oscillator (diatomic molecule, for instance) and the lowest Transverse Electric (TE) mode in a rectangular waveguide.  The QBIC effect will also occur in this waveguide system when the oscillator interacts with the second lowest TE mode, which has a cutoff frequency embedded in the continuum of the lowest TE mode.

\begin{figure}
\includegraphics[width=0.45\textwidth]{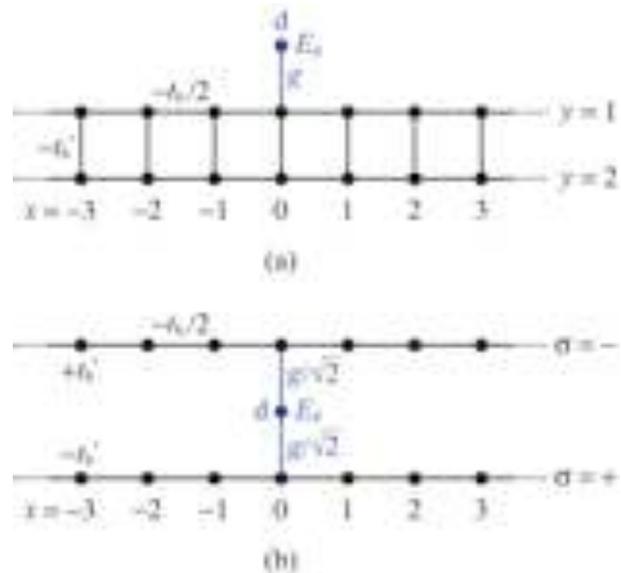}
\caption{(a) An adatom (quantum dot) attached to a ladder. (b) After partial diagonalization in the $y$ direction, the system is composed of the dot coupled to two independent channels.}
\label{fig-ladder}
\end{figure}

For the present case, we consider the system shown in Fig.~\ref{fig-ladder}(a), which is composed of two tight-binding chains and an adatom or quantum dot.
The two tight-binding chains (labeled $y=1,2$ in Fig.~\ref{fig-ladder}(a)) both have internal hopping parameter  $-t_\textrm{h}/2$ (internal sites of both chains are labeled by integer $x$, where $|x| \le m$ with $N=2m+1$ and $N$ $(\gg 1)$ is the number of sites in either chain $y=1$ or 2).
  The two chains are then coupled together site-by-site with hopping parameter $- t^{\prime}_\textrm{h}$, creating a ladder shape.  The dot (labeled d) is then coupled to the $x=0$ site of the $y=1$ chain.  Hence, we can write the Hamiltonian for our system as
\begin{eqnarray}
\hat{ \mathcal H} = 
& & -\frac{t_\textrm{h}}{2} \sum_{y = 1,2} \sum_{x} \left(  |x+1,y \rangle\langle x, y |+ |x, y \rangle\langle x+1, y | \right) \nonumber \\
& & - t^{\prime}_\textrm{h} \sum_{x} \left( |x,1 \rangle\langle x,  2 |+ |x, 2 \rangle\langle x, 1 | \right) \nonumber \\
& & + g \left( | \textrm d \rangle\langle 0, 1 |+ |0, 1 \rangle\langle \textrm d |  \right)  \nonumber \\  
& & + E_{\textrm d} | \textrm d \rangle\langle \textrm d | ,   \label{ladder.ham}
\end{eqnarray}
in which $E_{\textrm d}$ denotes the energy of the dot.  In accordance with our designations in Fig.~\ref{fig-ladder}, the first term here represents internal
hopping along either of the chains (in the $x$ direction) while the second term describes hopping from one chain to the other ($y=1$ to $y=2$ and vice versa).  The third term then represents hopping between the ad-atom and the $(0,1)$ site of the ladder and finally the fourth term gives the unshifted energy of the ad-atom.

In order to diagonalize the second term of the Hamiltonian~(\ref{ladder.ham}), we introduce the basis
\begin{equation}
\begin{pmatrix}
|x,+ \rangle \\
|x,- \rangle 
\end{pmatrix} 
\equiv 
U 
\begin{pmatrix}
|x,1 \rangle \\
|x,2 \rangle 
\end{pmatrix} ,
\label{eq.11}
\end{equation}
where 
\begin{equation}
U \equiv  \frac{1}{\sqrt{2}} \left(
\begin{array}{cc}
1 &  1 \\
1 &  -1 \\
\end{array} 
\right) = U^{-1}. 
\label{eq.12}
\end{equation}
Using the new basis $|x,\sigma=\pm \rangle$, the Hamiltonian can be divided into 
the $\sigma=+ $ and $-$~chains (with the adatom term) as
\begin{eqnarray}
\hat{ \mathcal H} = 
& & 
\sum_{\sigma=\pm}
\left[ 
-\frac{t_\textrm{h}}{2}  \sum_{x} \left(  |x+1,\sigma \rangle\langle x, \sigma |+ |x, \sigma \rangle\langle x+1, \sigma | \right)  \right. \nonumber \\
& & \left.  - \sigma t^{\prime}_\textrm{h} \sum_{x}  |x,\sigma \rangle\langle x,  \sigma |  + \frac{g}{\sqrt{2} } \left( | \textrm d \rangle\langle 0, \sigma |+ |0, \sigma \rangle\langle \textrm d |  \right)
\right] \nonumber \\ 
& & + E_\textrm{d} | \textrm d \rangle\langle \textrm d | .  \label{two-channel.ham}  
\end{eqnarray}
We have now obtained the Hamiltonian in the form of Fig.~\ref{fig-ladder}(b), in which the two  $\sigma=+,- $ chains represent two independent channels for charge transfer.
Note that we can also interpret the label $\sigma$ as electron spin and $t^{\prime}_\textrm{h}$ as a magnetic field.

In the next section, we will outline two approaches to obtaining the full diagonalization of the above Hamiltonian.  In the first approach, we will introduce the wave vector representation to obtain a bi-linear form of the Hamiltonian, from which we can obtain the energy eigenvalues of the system using the analysis due to Friedrichs~\cite{48Friedrichs}.  In the second approach we will rely on the recently presented method of outgoing waves~\cite{Hatano07}.

In Sec.~III we will present the full energy eigenvalue spectrum for the case in which the energy bands associated with the two channels of the quantum wire overlap; we show the energy shift and decay rate of the QBIC states.  
In particular, we will show that the decay rate of the QBIC states is proportional to $g^6$ and that this effect is a direct result of the interaction between the van Hove singularity at the edge of one energy band and the continuum of the other energy band.  We will also examine the wave function and time evolution for the QBIC state.  

In Sec.~IV we will examine the energy spectrum in two special cases.  In the first case the two channels become decoupled (that is, $t^{\prime}_\textrm{h} = 0$). The energy spectrum then reduces to that of a single channel quantum wire coupled with an adatom~\cite{06TGP}.
In the second case the lower edge of the upper band coincides with the upper edge of the lower band (that is, $t^{\prime}_\textrm{h} =  t_\textrm{h}$).  We will then examine the energy spectrum and discover a modification of the QBIC states resulting from the two overlapping singularities.

Finally, in Sec.~V we will briefly outline our results and make our final conclusions.
We will also discuss the generalization to an $n$-channel model briefly.
In Appendix, we summarize a numerical method of following the time evolution of decaying resonant states.

\section{Dispersion Relation and Diagonalization of the Hamiltonian}

In this section we will outline two methods for diagonalizing the Hamiltonian~(\ref{two-channel.ham}).  In the first we will write the Hamiltonian in a bi-linear form, from which we can immediately obtain the eigenvalues following the method due to Friedrichs~\cite{48Friedrichs}.  These eigenvalues will be obtained in the form of discrete solutions to a dispersion relation that is equivalent to a 12th order polynomial.  The discrete solutions of this polynomial give the diagonalized energy shifts and decay rates for an electron in the adatom.

In the second approach we will rely on the method of outgoing waves presented recently by N.~H., H.~N., K.~Sasada, and T.~P.~\cite{Hatano07}.  This method will also yield the dispersion relation, but is better suited for performing numerical simulations.

\subsection{Dispersion relation from the Friedrichs solution for bi-linear Hamiltonian}

\begin{figure}
\includegraphics[width=0.45\textwidth]{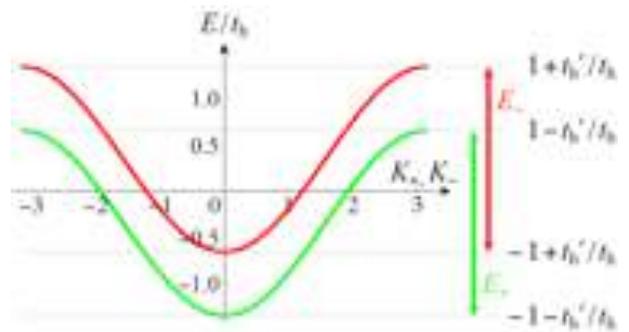}
\caption{The two continuous dispersions relations~(\ref{contDisp}) which form two conduction bands (channels) $E_\pm$ in the wire system. Here we graph the overlapping case in which $0<t^{\prime}_\textrm{h} < t_\textrm{h}$.}
\label{fig-TwoChannels}
\end{figure}

Because we are interested in the case $N \gg 1$, we may impose boundary conditions for our convenience of the calculations.
Imposing here the usual periodic boundary conditions in the $x$ direction,
we may introduce the wave vector representation with wave vectors $K_{\pm}$ in the two respective channels  $\sigma = \pm$ by
\begin{equation}
| K_{\pm} \ket = \frac{1}{\sqrt{N}} \sum_x e^{iK_{\pm}x} | x, \pm \ket,
\label{waveVector}
\end{equation}
where $K_\pm \equiv n_\pm \Delta k$ with $\Delta k \equiv 2\pi/N$ and integers $n_\pm$.
This allows us to write the Hamiltonian~(\ref{two-channel.ham}) as a variation of the bi-linear Friedrichs-Fano model
\begin{eqnarray}
\hat{ \mathcal H} &= &
\sum_{\sigma = \pm} \sum_{K_{\sigma}}
\left[  
E_{\sigma} | K_{\sigma} \rangle \langle K_{\sigma} | + \frac{g}{\sqrt{2N}} 
\left(| \textrm d \rangle \langle K_{\sigma} |  + | K_{\sigma}  \rangle \langle \textrm d | \right)\right]
\nonumber \\
& & + E_\textrm{d} | \textrm{d} \rangle\langle \textrm{d}|.
\label{wave-vector.ham}  
\end{eqnarray}
The energies $E_{\sigma}$ in the two channels are determined by their respective wave numbers $K_{\sigma}$ according to
\begin{equation}
E_{\pm} = - t_\textrm{h} \cos K_{\pm} \mp t^{\prime}_\textrm{h}. 
\label{contDisp}
\end{equation}
We refer to the above equations as the continuous dispersion equations for the system; this is because in the continuous limit $N \rightarrow \infty$ they describe the allowed energies for the two continua of $K_\sigma$ states.  In Fig.~\ref{fig-TwoChannels} we graph these two energy bands for the case $t^{\prime}_\textrm{h} < t_\textrm{h}$, in which they will overlap.  
(By contrast, below we will write a discrete dispersion equation 
that describes how the discrete energy $E_\textrm d$ is modified by the interaction.)
Both of the continuous channels described in Eq.~(\ref{contDisp}) have an associated density of states (DOS) function.  The DOS functions for the two channels are given by
\begin{equation}
\rho_{\pm}(E) = {1 \over \pi} {1 \over \sqrt{t_\textrm{h}^2 - (E \pm t^{\prime}_\textrm{h})^2 } }.
\label{DOS}
\end{equation}
Note the presence of two van Hove singularities in either channel.  These singularities are located at $ \pm t_\textrm{h} + t^{\prime}_\textrm{h}$ for the ``$+$"~channel and $ \pm t_\textrm{h} - t^{\prime}_\textrm{h}$ for the ``$-$"~channel.

Since the Hamiltonian~(\ref{wave-vector.ham}) is in a bi-linear form, in principle we may now diagonalize to explicitly solve the problem according to the method given by Friedrichs~\cite{48Friedrichs}.  However, we will leave the issue of obtaining the explicit solutions to the following subsection, in which the method is more suited to conducting numerical simulations for the time evolution of the system.  Instead, we will simply follow the standard method to obtain the discrete dispersion relation for an electron inside the two-channel quantum wire.  

The discrete dispersion relation is given by $\eta (z) \equiv z  - E_\textrm{d} - \Xi (z) = 0$, where the self-energy $ \Xi (z)$ for an electron in the two-channel wire is determined by
\begin{eqnarray}
\lefteqn{\Xi (z)\equiv  
\frac{g^2}{2N} \sum_{\sigma = \pm} \sum_{k_{\sigma}} \frac{1}{ z - E_{k_{\sigma} }} 
}
\nonumber \\
& = & 
 \frac{g^2}{2N} \left[ \sum_{k_+}  \frac{1}{z +  t^{\prime}_\textrm{h} - t_\textrm{h}\cos k_+}\right.
 \nonumber\\
 &&\phantom{ \frac{g^2}{2N} }\left.+ \sum_{k_-} \frac{1}{z - t^{\prime}_\textrm{h} - t_\textrm{h}\cos k_-} \right] \nonumber \\
&\stackrel{N\to \infty}{\longrightarrow}&
\frac{g^2}{4 \pi} \int_{-\pi}^{\pi} dk \left[ \frac{1}{z + t^{\prime}_\textrm{h} - t_\textrm{h} \cos k_+} \right.
\nonumber\\
&&
\phantom{\frac{g^2}{4 \pi} \int_{-\pi}^{\pi} dk}\left.+ \frac{1}{z - t^{\prime}_\textrm{h} - t_\textrm{h} \cos k_-} \right]  \nonumber \\
& = &
 \frac{g^2}{2} \left[ \frac{1}{\sqrt{(z + t^{\prime}_\textrm{h})^2 - t_\textrm{h}^2}}  + \frac{1}{\sqrt{(z - t^{\prime}_\textrm{h})^2 - t_\textrm{h}^2}} \right].\quad
\end{eqnarray}
Thus we find the dispersion equation
\begin{equation}
z - E_\textrm{d} - \frac{g^2}{2} \left[ \frac{1}{\sqrt{(z + t^{\prime}_\textrm{h})^2 - t_\textrm{h}^2}}  + \frac{1}{\sqrt{(z - t^{\prime}_\textrm{h})^2 - t_\textrm{h}^2}} \right] = 0
\label{adatomDisp}
\end{equation}
as reported previously~\cite{QBIC-Letter,SG-Diss}.  This dispersion equation describes the behavior of an electron initially trapped in the adatom 
and can be written as a 12th order polynomial equation in $z$.  Hence, we will also refer to this equivalent equation as the {\it dispersion polynomial}.  The twelve discrete solutions to this equation give the allowed bound states (purely real solutions) and resonant states (complex solutions) in the diagonalized system.  As we will discuss below, these solutions can be viewed as living in a complex energy surface, parameterized by the original impurity energy $E_{\textrm d}$.  In the present case of the two-channel wire, this energy surface will be composed of four Riemann sheets.

Once we have obtained the twelve discrete solutions $z = E$ to the dispersion polynomial, then Eq.~(\ref{contDisp}) implies that each eigenvalue $E$ can be assigned two $K_{\pm}$ values.  With the values $K_{\pm}$ in hand, we will be able to write the wave function for each solution, making use of Eq.~(\ref{eq.14}) given below.  Practically, this means that any electron state in the wire can be fully described by the three values $E$, $K_+$ and $K_-$.


\subsection{Solutions of Hamiltonian by the method of two outgoing waves}

We will now solve the Schr\"odinger equation
\begin{equation}
\hat{ \mathcal H} | \psi \rangle= E  | \psi \rangle,
\label{Sch-eqn}
\end{equation}
for the resonant states $| \psi \rangle$ of the Hamiltonian given in Eq.~(\ref{two-channel.ham}).  As has been previously shown~\cite{Hatano07,05Sasada},
 the resonant eigenfunction of the tight binding model on a chain with an adatom can be written in the form
\begin{equation}
\psi_{\sigma} (x) \equiv \langle x,\sigma | \psi  \rangle = A_{\sigma} e^{i K_\sigma |x|}, \label{eq.14}
\end{equation}
or
\begin{equation}
\vec\psi_{\sigma} (x) \equiv 
\begin{pmatrix}
\psi_{+} (x) \\
\psi_{-} (x) 
 \end{pmatrix} 
 = A_{+} e^{iK_{+} |x| } 
\begin{pmatrix}
1 \\
0
\end{pmatrix} 
+ A_{-} e^{iK_{-} |x| } 
\begin{pmatrix}
0 \\
1
\end{pmatrix}  
. \label{eq.15}
\end{equation}
Using the resonant eigenfunction~(\ref{eq.14}) in the Schr\"odinger equation~(\ref{Sch-eqn})
for the case $x\ne0$, we obtain
\begin{eqnarray}
E \psi_{\sigma} (x) &=&
\hat{ \mathcal H} \  \psi_{\sigma} (x)  \nonumber \\ 
&=& -\frac{t_\textrm{h}}{2} \left\{ \psi_{\sigma} (x+1) +\psi_{\sigma} (x-1) \right\} 
-\sigma  t^{\prime}_\textrm{h} \psi_{\sigma} (x)  \nonumber \\
&=& \left( - t_\textrm{h} \cos K_{\sigma} - \sigma t^{\prime}_\textrm{h} \right) \psi_{\sigma} (x). \label{eq.16}
\end{eqnarray}
This again yields the dispersion relations $E= - t_\textrm{h} \cos K_{\pm} \mp t^{\prime}_\textrm{h}$ inside the two channels of the wire as given in Eq.~(\ref{contDisp}) above.

To solve the eigenequation 
\begin{equation}
\psi_{\textrm d}  \equiv   \langle \textrm{d}|\psi \rangle \quad \textrm{and} \quad
{\psi} (x,y) \equiv  \langle x,y |\psi \rangle, 
\label{eq.30} 
\end{equation}
for $x=0$ and $x=\textrm{d}$, respectively, we return to the original $y$ space.
Using the original bases $|x,y \rangle$ and Eq.~(\ref{eq.11}), the resonant eigenfunctions 
$\psi(x,y) $ in the original $y$ space have the form
\begin{eqnarray}
\vec{\psi}_{y} (x) &\equiv &
\begin{pmatrix}
{\psi} (x,1)\\
{\psi} (x,2)
\end{pmatrix} 
\nonumber \\
&=& U \vec{\psi}_{\sigma} (x) \nonumber \\
&=&  \frac{1}{\sqrt{2}} A_{+} e^{iK_{+} |x| } 
\begin{pmatrix}
1 \\
1
\end{pmatrix} 
+ \frac{1}{\sqrt{2}} A_{-} e^{iK_{-} |x| } 
\begin{pmatrix}
1 \\
-1
\end{pmatrix}.
\nonumber \\
& & \label{eq.40}
\end{eqnarray}

With the Hamiltonian~(\ref{ladder.ham}) the Schr\"odinger equations~(\ref{Sch-eqn}) for $x = 0$ and $x = \textrm{d}$ become
\begin{equation}
\begin{cases}
\displaystyle -\frac{t_\textrm{h}}{2} \left\{ {\psi} (-1,1) + {\psi} (1,1)  \right\} - t^{\prime}_\textrm{h} {\psi} (0,2) + g {\psi}_{\textrm d} \nonumber \\
\hspace{6cm} = E {\psi} (0,1) ,  \\
\displaystyle -\frac{t_\textrm{h}}{2} \left\{ {\psi} (-1,2) + {\psi} (1,2)  \right\} - t^{\prime}_\textrm{h} {\psi} (0,1)  \nonumber \\
\hspace{6cm} = E {\psi} (0,2) , \\
g {\psi} (0,1) +E_{\textrm d}  {\psi}_{\textrm d} 
= E {\psi}_{\textrm d}. 
\end{cases}
\\ \label{eq.41} 
\end{equation}
Substituting the resonant wave functions $\psi$ in the site representation from Eq.~(\ref{eq.40}) while making use of the continuous dispersion relations Eq.~(\ref{contDisp}), 
we obtain
\begin{equation}
\begin{cases}
i t_\textrm{h} A_{+} \sin K_{+}  + i t_\textrm{h} A_{-} \sin K_{-}  - \sqrt{2} g {\psi}_{\textrm d} = 0 ,  \\
i t_\textrm{h} A_{+} \sin K_{+}  -  i t_\textrm{h} A_{-} \sin K_{-} = 0 , \\
g \left( A_{+} + A_{-} \right) + \sqrt{2} \left( E_{\textrm d} - E \right)  {\psi}_{\textrm d} = 0, \end{cases}
\\  \label{eq.46}
\end{equation}
which can be written in matrix form as
\begin{equation}
\left(
\begin{array}{ccc}
it_\textrm{h}\sin K_{+} &  it_\textrm{h} \sin K_{-} & -\sqrt{2}g  \\
it_\textrm{h}\sin K_{+} & -it_\textrm{h} \sin K_{-} & 0  \\
g &  g  & \sqrt{2} \left( E_{\textrm d} - E \right)   
\end{array}
\right)
\begin{pmatrix}
A_{+} \\
A_{-} \\
{\psi}_{\textrm d}
\end{pmatrix}
= 0. \label{eq.47}
\end{equation}

In order to have non-trivial solutions to Eq.~(\ref{eq.47}), the determinant of the coefficient matrix above must be zero.  Hence we obtain the following condition on $A_{+},~A_{-}$ and $\psi_{\textrm d}$:
\begin{equation}
E-E_{\textrm d} = g^2
\left(
\frac{1}{2it_\textrm{h} \sin K_{+}} + 
\frac{1}{2it_\textrm{h} \sin K_{-}} 
\right). \label{eq.49}
\end{equation}
Making use of the continuous dispersion equations~(\ref{contDisp}), we can see that the above condition is equivalent to the discrete dispersion equation~(\ref{adatomDisp}) related to the 
interaction between the adatom and the two $\sigma$ channels.

According to the previous work~\cite{06TGP,Hatano07}, the dispersion equation for the single-chain model  with an adatom is given by 
\begin{equation}
E_{\textrm{chain}}-E_{\textrm d} =\frac{g^2}{ 2 it_\textrm{h} \sin K_{\textrm{chain}} }
\end{equation}  
Hence we note that the present dispersion equation~(\ref{eq.49}) for the ladder model corresponds to the sum of two single-chain dispersion equations.



In the case of the single chain model~\cite{06TGP,SG-Diss,Hatano07}, the complex energy spectrum could be evaluated in terms of a complex plane consisting of two Riemann sheets.  In that case there was only one wave number $K_{\textrm{chain}}$ corresponding to the single channel available to an electron.  One can then easily classify whether a resonant state lies in the first or second Riemann sheet according to the sign of the imaginary component of this wave number.  
This determination is beneficial in obtaining a detailed understanding of the survival probability for the excited state, in which a contour deformation must be performed in the complex energy plane.
For instance, the position of the poles may also influence the strength of the non-Markovian effect due to the so-called branch point effect (this will be the subject of a future publication).

\begin{figure}
\includegraphics[width=0.45\textwidth]{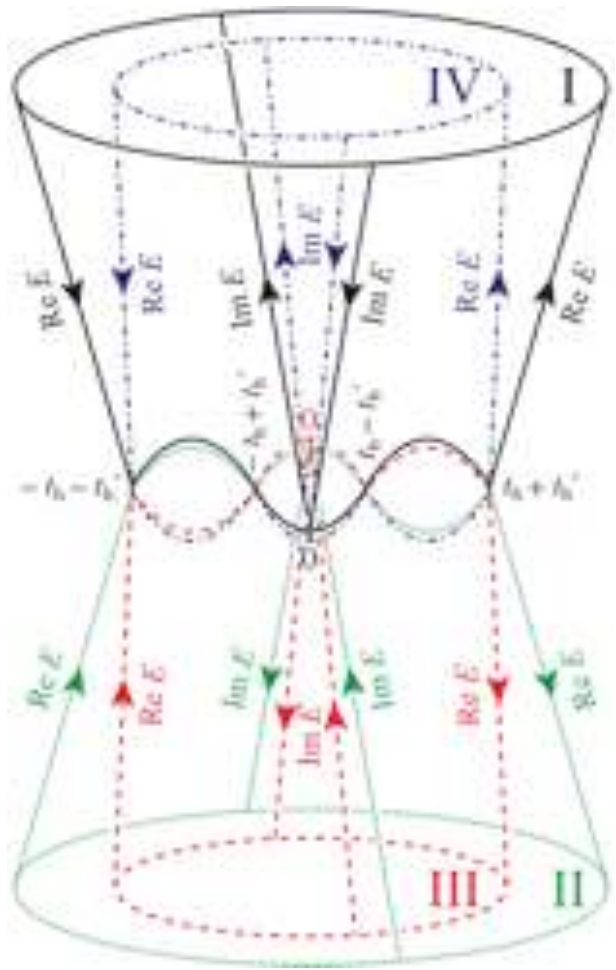}
\caption{
Four sheeted Riemann energy surface for the two channel model in the case $0 < t_\textrm{h}^\prime < t_\textrm{h}$. 
 Solutions of the discrete dispersion equation with the sign combination $(\mathrm{sgn} (\mathrm{Im} K_+), \mathrm{sgn} (\mathrm{Im} K_-)) = (+,+)$ lie in the first Riemann sheet (solid black line).  Solutions with the combination $(-,+)$ lie in Sheet~II (short-dashed green line), those with the combination $(+,-)$ line in Sheet~III (long-dashed red line), and those with the combination $(-,-)$ lie in Sheet~IV (chained blue line).  The curved lines in the center of the diagram represent the two branch cuts where the sheets intersect along their respective real axes.  For instance, on the left side of the diagram, if one starts from the positive imaginary half of Sheet~I and crosses the real axis between $- t_\textrm{h} -t^{\prime}_\textrm{h}$ and $- t_\textrm{h}+t^{\prime}_\textrm{h}$ (represented by the curved overlapping solid black and short-dashed green lines) then one will emerge on the negative imaginary half of Sheet~II.
}
\label{FIG-Riemann}
\end{figure}

In the present case of the two-channel model, there are two wave numbers $K_{\pm}$ resulting from the two channels available to the electron.  The imaginary part of these two wave numbers together provides four possible sign combinations and hence the complex energy plane is now composed of four Riemann sheets; see Fig.~\ref{FIG-Riemann} for an example in the case $0 < t_\textrm{h}^\prime < t_\textrm{h}$.
We can also see that the Riemann surface must be four-sheeted by considering the dispersion equation~(\ref{adatomDisp}), in which each of the two roots may take either a positive or negative sign, again resulting in four combinations (although one must be careful here as the sign combinations in this approach change for different portions of the same sheet).  In the general case of an $n$-channel quantum wire, the complex energy surface will be composed of $2n$ Riemann sheets.  We will comment further on the generalization to an arbitrary number of channels later in this paper.

For the purpose of assignment of each solution to the correct Riemann sheet, it is more convenient to modify Eq.~(\ref{eq.49}) with the help of the continuous dispersion equation~(\ref{contDisp}) and solve the following set of simultaneous equations with respect to $K_\pm$ than to solve the discrete dispersion equation~(\ref{adatomDisp}) with respect to $z=E$:
\begin{eqnarray}
\lefteqn{
-t_\textrm{h} \cos K_+ -t^{\prime}_\textrm{h}
=-t_\textrm{h} \cos K_- +t^{\prime}_\textrm{h}
}
\nonumber
\\
& = &E_{\textrm d} +g^2
\left(
\frac{1}{2it_\textrm{h} \sin K_{+}} + 
\frac{1}{2it_\textrm{h} \sin K_{-}} 
\right). \label{eq.49-1}
\end{eqnarray}
The signs of $\mathop{\mathrm{Im}} K_\pm$ of each solution give the correct Riemann sheet immediately.

Note that the first Riemann sheet is assigned in a natural way as the energy eigenvalues of all of the solutions in this sheet must be real, each corresponding to a bound state in the energy spectrum analysis.
(They are on the positive imaginary axes of the wave-number spaces $K_\pm$.)
This is because the Hamiltonian must behave in a manner equivalent to a Hermitian operator in this first sheet with all eigenvalues being purely real.
It is only when the Hamiltonian is extended into the rigged Hilbert space~\cite{Berggren}
that complex solutions in the other sheets may be considered on an equal footing with the stable solutions in the first sheet, and that they can be interpreted as complex eigenvalues of the Hamiltonian in the rigged Hilbert space~\cite{PPT91,AP93}.


\section{Energy spectrum analysis for $0 < t_\textrm{h}^\prime < t_\textrm{h}$ case and QBIC states}

\begin{figure}
\includegraphics[width=0.45\textwidth]{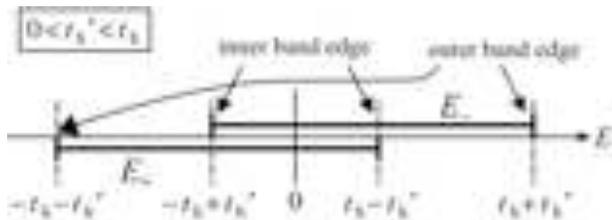}
\caption{Band structure for the case $0 < t_\textrm{h}^\prime < t_\textrm{h}$. See Eq.~(\ref{contDisp}) for the definition of $E_\pm$.
}
\label{FIG-bands}
\end{figure}

In this section we analyze the eigenenergy spectrum for the case $0 < t_\textrm{h}^\prime < t_\textrm{h}$, in which the two conduction bands overlap.  In this overlapping case the edge of one band is embedded in the continuum of the other and vice-versa.  This results in two outer band edges and two inner (embedded) band edges (see Fig.~\ref{FIG-bands}).

When the energy of the impurity lies inside the overlapping region between the two inner band edges ($-t_\textrm{h}+t^\prime_\textrm{h} < E_\textrm{d} < t_\textrm{h}-t^\prime_\textrm{h}$), we will find that there exist persistent stable solutions that lie just outside of the two outer band edges, as mentioned in Introduction~\cite{06TGP,Mahan,05PTG,SG-Diss}.  As we will see, these stable states are a direct result of the van Hove singularity in the density of states at the band edge.  For the inner band edges, however, there will now be two competing effects: the first being the stabilizing effect of the singularity from the embedded band edge and the second being the de-stabilizing effect of the continuum in which it is embedded (i.e., the second conduction band).  We will see that this will result in a slightly de-stabilized state embedded in the continuum (the QBIC state).

\subsection{Energy spectrum analysis from {\bf the} dispersion polynomial}

\begin{figure*}
\includegraphics[width=0.90\textwidth]{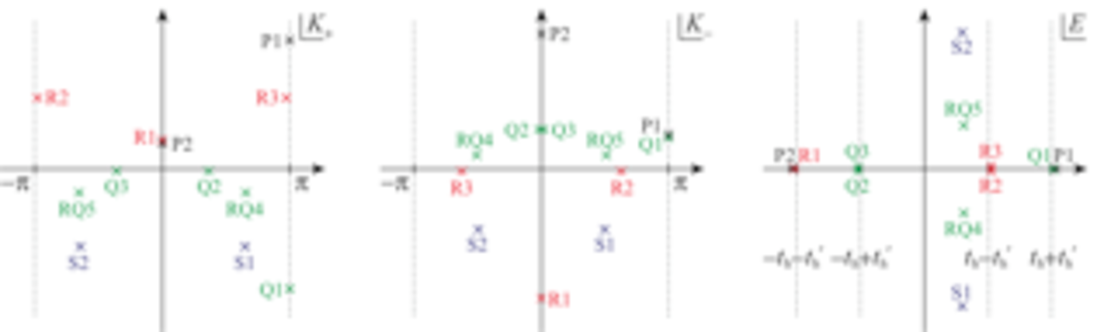}
\\
\hspace*{0.17\textwidth}(a)\hfill(b)\hfill(c)\hspace*{0.175\textwidth}
\caption{The relative positions of the poles in the complex $K_+$, $K_-$ and $E$ planes for the value  $E_\textrm{d} = 0.3$, as well as $t^\prime_\textrm{h} = 0.345$ and $g = 0.1$, as given in Table~\ref{table.pole}.
The two vertical dotted lines in the figures~(a) and~(b) indicate the edges of the Brillouin zone $\mathop{\mathrm{Re}}K_\pm=\pm\pi$, while the four vertical dotted lines in the figure~(c) represent the van Hove singularities.}
\label{FIG-eigenvalues}
\end{figure*}

We will now analyze in detail the twelve solutions to the discrete dispersion equation~(\ref{adatomDisp}).  Before presenting the full complex energy spectrum as a function of $E_\textrm{d}$, we will first consider the eigenenergies at two specific values of $E_\textrm{d}$ in Tables~\ref{table.pole} and~\ref{table2.pole}, in order to illustrate our earlier point regarding the placement of the solutions in the complex energy surface.
\begin{table*}
\caption{The twelve discrete eigenvalues for $t^\prime_\mathrm{h} = 0.345t_\mathrm{h}, g = 0.1t_\mathrm{h}$, and $E_\mathrm{d}=0.3t_\mathrm{h}$.  The decay solutions are indicated by shading.}
\label{table.pole}
\begin{ruledtabular}
\begin{tabular}{c||rl|rl|rl|c}
state & $E / t_\mathrm{h}$ & & $K_{+}$ & &$K_{-}$ & & Riemann Sheet \\
\hline 
\textrm{P1(0.3)} &      1.34501152 &                                           &       3.14159265   & $+i$ 1.11593256                   &      3.14159265  & $+i$ 0.00480148  & I \\
\textrm{P2(0.3)}& $-$1.34500463 &                                          &                          & $+i$ 0.00304629                   &                        & $+i$ 1.11592751  & I \\
\hline
\textrm{Q1(0.3)} &      1.34501136 &                                           &       3.14159265   & $-i$ 1.11593245                   &      3.14159265  & $+i$ 0.00476787  & II \\
\textrm{Q2(0.3)} & $-$0.65501370 &  $-i$ 1.5093  $\times 10^{-7}$ &       1.25558888   & $-i$ 1.5875  $\times 10^{-7}$ & $-$0.00002882  & $+i$ 0.00523534  & II \\ \shadecells{1-8}
\textrm{Q3(0.3)}  & $-$0.65501370 & $+i$ 1.5093  $\times 10^{-7}$ &  $-$1.25558888   & $-i$ 1.5875  $\times 10^{-7}$ &      0.00002882  & $+i$ 0.00523534  & II \\
\textrm{RQ4(0.3)} &     0.29998854 &   $-i$ 0.00153774                   &       2.27180290   & $-i$ 0.00201224                   & $-$1.52576970  & $+i$ 0.00153930  & II \\ \shadecells{1-8}
\textrm{RQ5(0.3)} &      0.29998854 &  $+i$ 0.00153774                   &  $-$2.27180290   & $-i$ 0.00201224                   &      1.52576970  & $+i$ 0.00153930  & II \\
\hline
\textrm{R1(0.3)}  & $-$1.34500459 &                                           &                         & $+i$ 0.00303273                   &                        & $-i$ 1.11592748  & III \\
\textrm{R2(0.3)} &      0.65509906 &  $-i$ 2.9331  $\times 10^{-6}$ &  $-$3.14138429   & $+i$ 0.01407702                   &      1.88609355  & $-i$ 3.0852  $\times 10^{-6}$ & III \\ \shadecells{1-8}
\textrm{R3(0.3)}  &      0.65509906 & $+i$ 2.9331  $\times 10^{-6}$ &       3.14138429   & $+i$ 0.01407702                   & $-$1.88609355  & $-i$ 3.0852  $\times 10^{-6}$ & III \\
\hline
\textrm{S1(0.3)} &     0.29991927 &  $-i$ 0.01154476                  &       2.27161773   & $-i$ 0.01510419                   &      1.52570333  & $-i$ 0.01155625  & IV \\ \shadecells{1-8}
\textrm{S2(0.3)} &      0.29991927 &  $+i$ 0.01154476                   &  $-$2.27161773   & $-i$ 0.01510419                   & $-$1.52570333  & $-i$ 0.01155625  & IV 
\end{tabular}
\end{ruledtabular}
\end{table*}
\begin{table*}
\caption{The twelve discrete eigenvalues for $t^\prime_\mathrm{h} = 0.345t_\mathrm{h}, g = 0.1t_\mathrm{h}$, and $E_\mathrm{d}=-1.0t_\mathrm{h}$.  The decay solutions are indicated by shading.}
\label{table2.pole}
\begin{ruledtabular}
\begin{tabular}{c||rl|rl|rl|c}
state & $E / t_\mathrm{h}$ & & $K_{+}$ & &$K_{-}$ & & Riemann Sheet \\
\hline 
\textrm{P1($-$1)} &     1.34500228 &                             &     3.14159265 & $+i$  1.11592578   &     3.14159265 & $+i$ 0.00213553             & I \\
\textrm{P2($-$1)} & $-$ 1.34510721 &                             &                & $+i$  0.01464344   &                & $+i$ 1.11600280             & I \\
\hline
\textrm{Q1($-$1)}  &     1.34500226 &                             &     3.14159265 & $-i$ 1.11592577   &     3.14159265 & $+i$ 0.00212886             & II \\
\textrm{Q2($-$1)} &  $-$ 1.00545676 &  $-i$ 0.00659855            &     0.84940336 & $-i$ 0.00878757   & $-$ 0.00726999 & $+i$ 0.81454288             & II \\ \shadecells{1-8}
\textrm{Q3($-$1)}  & $-$ 1.00545676 & $+i$ 0.00659855             & $-$ 0.84940336 & $-i$ 0.00878757   &     0.00726999 & $+i$ 0.81454288             & II \\
\hline
\textrm{R1($-$1)} & $-$ 1.34510275 &                             &                & $+i$ 0.01433550   &                & $-i$ 1.11599952             & III \\
\textrm{R2($-$1)} &     0.65500456 &  $-i$ 2.9002$\times 10^{-8}$ & $-$ 3.14158305 & $+i$ 0.00302110   &     1.88599415 & $-i$ 3.0505$\times 10^{-8}$ & III \\ \shadecells{1-8}
\textrm{R3($-$1)} &     0.65500456 & $+i$ 2.9002$\times 10^{-8}$ &     3.14158305 & $+i$ 0.00302110   & $-$ 1.88599415 & $-i$ 3.0505$\times 10^{-8}$ & III \\
\textrm{RQ4($-$1)} &  $-$ 0.65510500 &  $+i$ 3.2049$\times 10^{-6}$ &  1.25549284 & $+i$ 3.3711$\times 10^{-6}$ & $-$  0.00022112 & $-i$ 0.01449328 & III \\ \textrm{RQ5($-$1)}  & $-$ 0.65510500 & $-i$ 3.2049$\times 10^{-6}$ & $-$  1.25549284 & $+i$ 3.3711$\times 10^{-6}$ &  0.00022112 & $-i$ 0.01449328 & III \\
\shadecells{1-8}
\hline
\textrm{S1($-$1)} &  $-$ 0.99434007 &  $-i$ 0.00663666             &     0.86411247 & $-i$ 0.00872637   &     0.00744838 & $-i$ 0.80218340             & IV \\ \shadecells{1-8}
\textrm{S2($-$1)}  & $-$ 0.99434007 & $+i$ 0.00663666             & $-$ 0.86411247 & $-i$ 0.00872637   & $-$ 0.00744838 & $-i$ 0.80218340             & IV 
\end{tabular}
\end{ruledtabular}
\end{table*}
In these tables, we present the values obtained for each of the twelve solutions by solving the dispersion equation~(\ref{adatomDisp}) for the eigenenergies $E$ and the continuous dispersion equations~(\ref{contDisp}) for the wave number pairs $K_{\pm}$ 
at the two values $E_\textrm{d} = 0.3t_\textrm{h}$ and $E_\textrm{d} = -1.0t_\textrm{h}$, respectively.
For the other parameters of the system, in both tables we have used the values $t^\prime_\textrm{h} = 0.345 t_\textrm{h}$ and $g = 0.1 t_\textrm{h}$.  In these tables (and throughout this paper) we measure energy in units $t_\textrm{h} = 1$.
In addition, in Fig.~\ref{FIG-eigenvalues} we have indicated the relative position of each individual pole in the complex $K_+$, $K_-$ and $E$ planes, respectively, each given for the value $E_\textrm{d} = 0.3 t_\textrm{h}$ from Table~\ref{table.pole}.

\begin{figure*}
\hspace*{0.05\textwidth}
\includegraphics[width=0.4\textwidth]{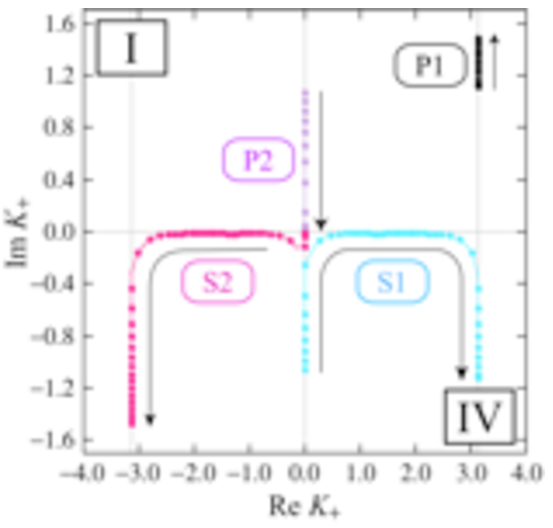}
\hfill
\includegraphics[width=0.4\textwidth]{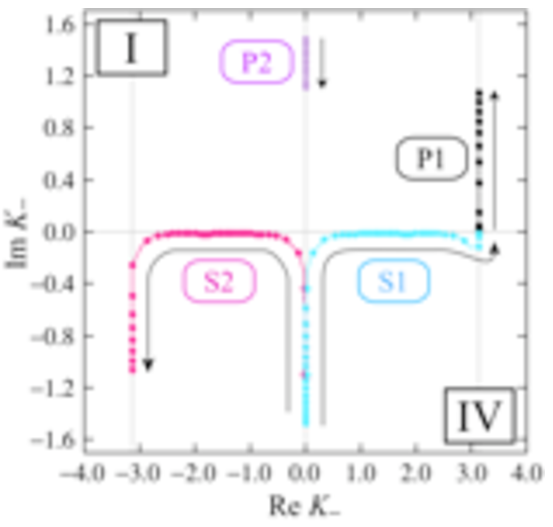}
\hspace*{0.05\textwidth}
\\
\vspace*{-\baselineskip}
\hspace*{0.08\textwidth}(a)\hspace*{0.465\textwidth}(b)\hspace*{0.4\textwidth}
\\
\vspace*{2\baselineskip}
\hspace*{0.05\textwidth}
\includegraphics[width=0.4\textwidth]{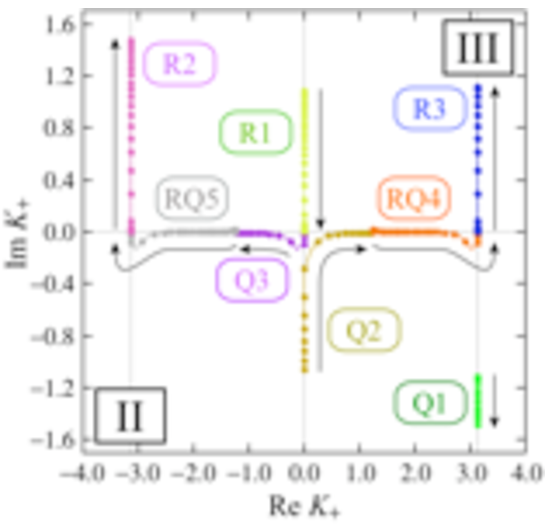}
\hfill
\includegraphics[width=0.4\textwidth]{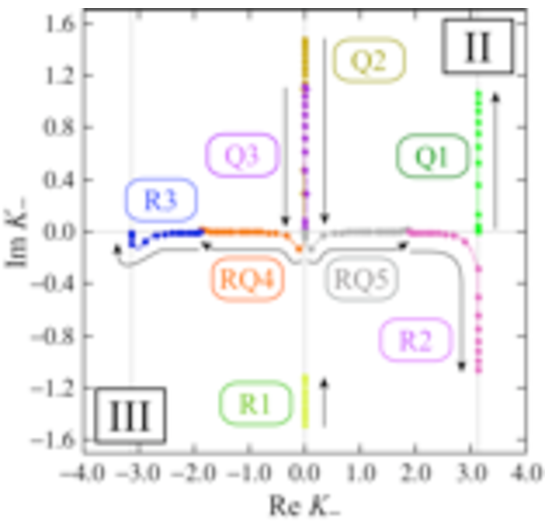}
\hspace*{0.05\textwidth}
\\
\vspace*{-\baselineskip}
\hspace*{0.08\textwidth}(c)\hspace*{0.465\textwidth}(d)\hspace*{0.4\textwidth}
\vspace*{\baselineskip}
\caption{The wave numbers $K_\pm$ for the twelve solutions of the set of equations~(\ref{eq.49-1}).
The arrows represent how the solutions move when we increase $E_\mathrm{d}$ from $-2t_\textrm{h}$ to $2t_\textrm{h}$ with $t^\prime_\textrm{h} = 0.345 t_\textrm{h}$ and $g = 0.1 t_\textrm{h}$.
The left-hand figures~(a) and~(c) show the $K_+$ plane and the right-hand figures~(b) and~(d) show the $K_-$ plane.
In the top figures~(a) and (b), the solutions in the upper half plane are in the Riemann sheet~I and those in the lower half plane are in the Riemann sheet~IV.
In the figure~(c), the solutions in the lower half plane are in the Riemann sheet~II and those in the upper half plane are in the Riemann sheet~III.
In the figure~(d), the solutions in the upper half plane are in the Riemann sheet~II and those in the lower half plane are in the Riemann sheet=III.
In the bottom figures~(c) and~(d), the solutions RQ4 and RQ5 cross the real axes from the Riemann sheet~III into the Riemann sheet~II when $E_\textrm{d}$ is increased from negative to positive.
The vertical gray lines represent $K_\pm=-\pi,0,\pi$.
}
\label{fig-K+-}
\end{figure*}

\begin{figure*}
\hspace*{0.05\textwidth}
\includegraphics[width=0.4\textwidth]{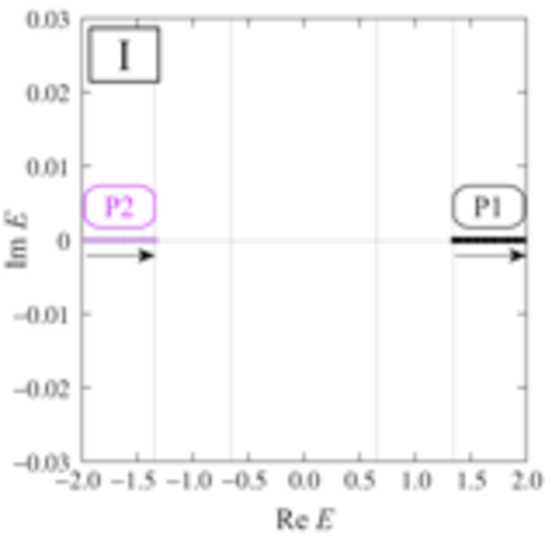}
\hfill
\includegraphics[width=0.4\textwidth]{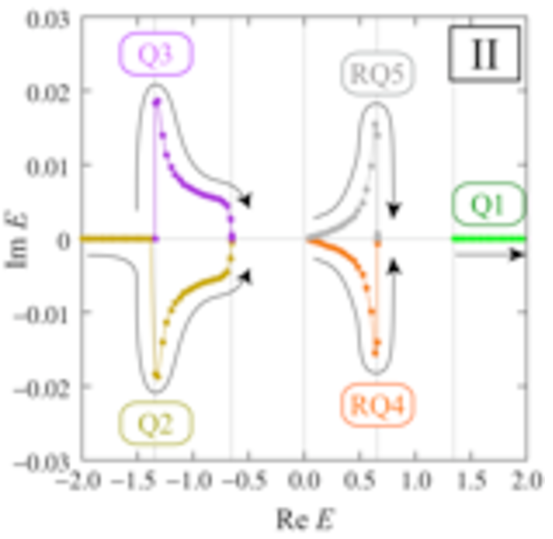}
\hspace*{0.05\textwidth}
\\
\vspace*{-\baselineskip}
\hspace*{0.08\textwidth}(a)\hspace*{0.465\textwidth}(b)\hspace*{0.4\textwidth}
\\
\vspace*{2\baselineskip}
\hspace*{0.05\textwidth}
\includegraphics[width=0.4\textwidth]{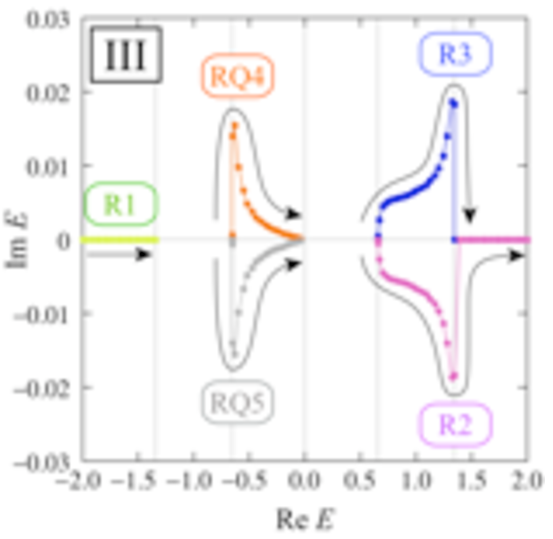}
\hfill
\includegraphics[width=0.4\textwidth]{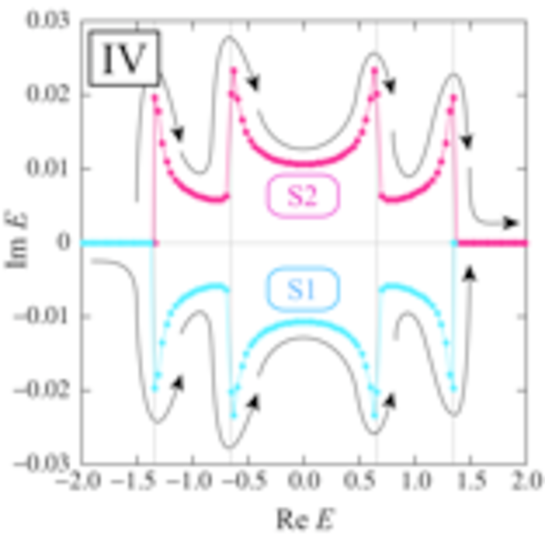}
\hspace*{0.05\textwidth}
\\
\vspace*{-\baselineskip}
\hspace*{0.08\textwidth}(c)\hspace*{0.465\textwidth}(d)\hspace*{0.4\textwidth}
\vspace*{\baselineskip}
\caption{The energy for the twelve solutions of the dispersion equation, put in the corresponding Riemann sheets.
The arrows represent how the solutions move when we increase $E_\mathrm{d}$ from $-2t_\textrm{h}$ to $2t_\textrm{h}$ with $t^\prime_\textrm{h} = 0.345 t_\textrm{h}$ and $g = 0.1 t_\textrm{h}$.
Energy is measured in units of $t_\textrm{h}=1.$
The vertical gray lines represent the four van Hove singularities.
}
\label{fig-E}
\end{figure*}

As mentioned previously, the placement of each solution in the complex energy plane can be determined in a straight-forward manner by the sign of the imaginary parts of the two complex wave vectors $K_{\pm}$,
which are given as the solutions of the simultaneous equations~(\ref{eq.49-1});
see Fig.~\ref{fig-K+-}.
We have designated those with positive imaginary $K_+$ component and positive imaginary $K_-$ component ($(+,+)$, respectively) as lying in Riemann Sheet~I; likewise we have designated $(-,+)$ as Sheet~II, $(+,-)$ as Sheet~III and $(-,-)$ as Sheet~IV.
The resulting eigenenergy $E=-t_\textrm{h}\cos K_\pm \mp t^\prime_\textrm{h}$ of each solution is shown in Fig.~\ref{fig-E}.
In Tables~\ref{table.pole} and~\ref{table2.pole} and Figs.~\ref{fig-K+-} and~\ref{fig-E}, we have labeled each solution by a letter P, Q, R or S according to the Riemann sheet (I, II, III, or IV, respectively) in which that solution lies.
  
As these solutions are roots of a polynomial with real coefficients, each complex decay solution (with negative imaginary part) is accompanied by a complex conjugate ``growth'' solution with positive imaginary part.  However, when we calculate the survival probability for the excited impurity state, only the decay solutions will contribute a pole in the complex contour integration.  Hence, our focus will be the decay solutions indicated by shades in Tables~\ref{table.pole} and~\ref{table2.pole}.
 
Note that Sheet~I contains only purely real solutions.  This choice is necessary as our Hamiltonian should be defined to reduce to a Hermitian operator (with real eigenvalues) in the first sheet in the finite case (closed system) or when the coupling vanishes.  
Also notice that there is one conjugate pair of solutions (RQ4 and RQ5) labeled by two sheets.  This is due to the fact that these solutions lie in Sheet~III for negative values of $E_\textrm{d}$ then cross into Sheet~II (through the double branch cut on the real axis of the complex energy plane) for positive values of  $E_\textrm{d}$.  
At the same time, the imaginary part of the energy eigenvalue changes its sign;
the solution RQ5 is the decay solution for $E_\textrm{d}<0$, but the solution RQ4 is the one for $E_\textrm{d}>0$.
These solutions disappear for the case $ t_\textrm{h}= t^\prime_\textrm{h}$ as will be discussed in Sec.~IV.



\begin{figure*}
\hspace*{0.05\textwidth}
\includegraphics[width=0.4\textwidth]{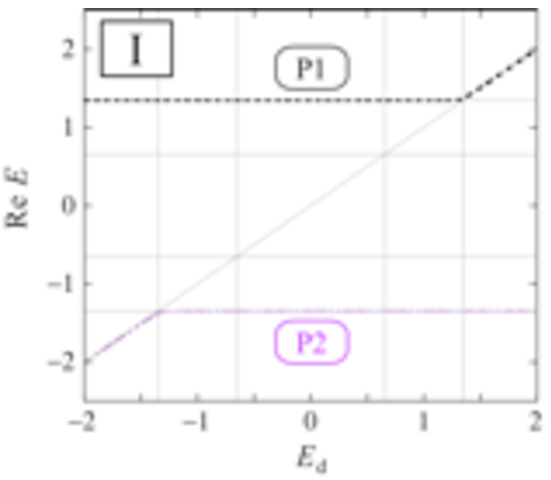}
\hfill
\includegraphics[width=0.4\textwidth]{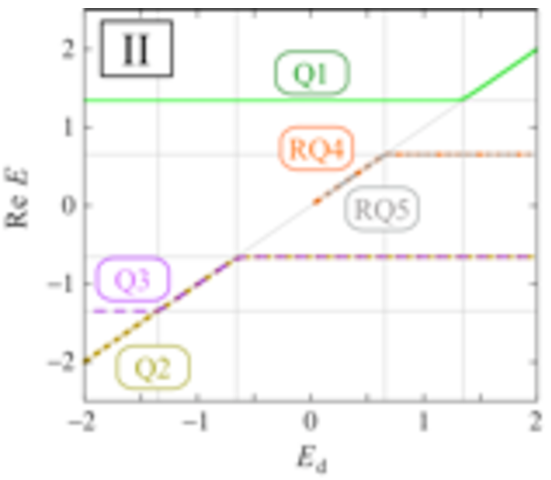}
\hspace*{0.05\textwidth}
\\
\vspace*{-\baselineskip}
\hspace*{0.08\textwidth}(a)\hspace*{0.465\textwidth}(b)\hspace*{0.4\textwidth}
\\
\vspace*{2\baselineskip}
\hspace*{0.05\textwidth}
\includegraphics[width=0.4\textwidth]{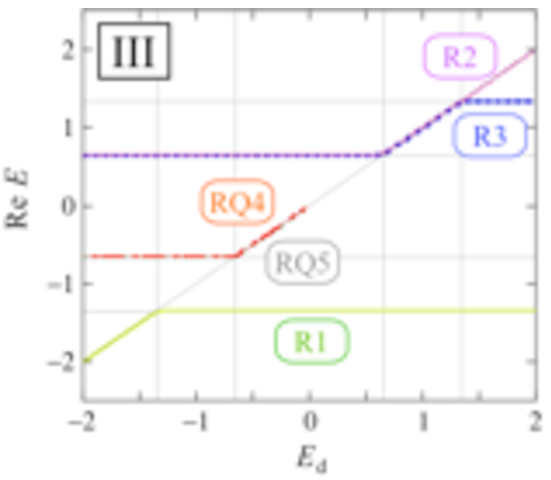}
\hfill
\includegraphics[width=0.4\textwidth]{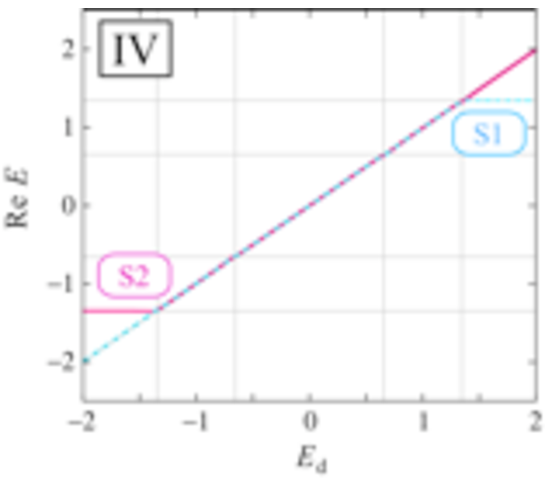}
\hspace*{0.05\textwidth}
\\
\vspace*{-\baselineskip}
\hspace*{0.08\textwidth}(c)\hspace*{0.465\textwidth}(d)\hspace*{0.4\textwidth}
\\
\vspace*{2\baselineskip}
\hspace*{0.05\textwidth}
\includegraphics[width=0.4\textwidth]{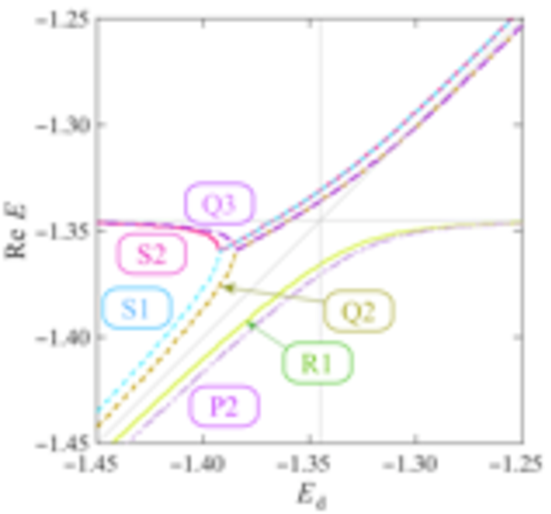}
\hfill
\includegraphics[width=0.4\textwidth]{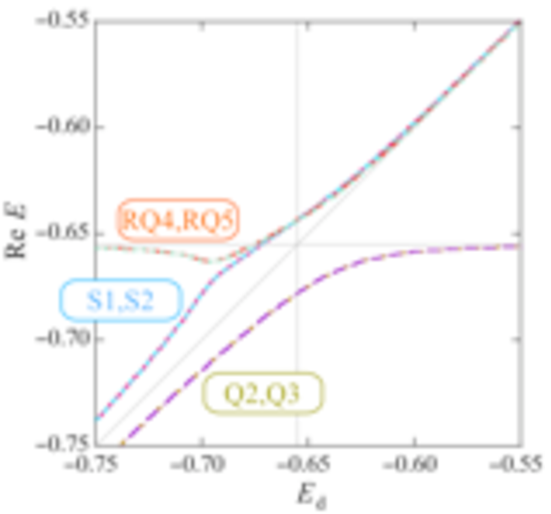}
\hspace*{0.05\textwidth}
\\
\vspace*{-\baselineskip}
\hspace*{0.08\textwidth}(e)\hspace*{0.465\textwidth}(f)\hspace*{0.4\textwidth}
\vspace*{\baselineskip}
\caption{Real part of the energy for the twelve solutions of the dispersion equation as a function of $E_\textrm{d}$ for the values $t^\prime_\textrm{h} = 0.345 t_\textrm{h}$ and $g = 0.1 t_\textrm{h}$.
Energy is measured in units of $t_\textrm{h}=1.$
In the top four figures~(a)--(d), each solution is plotted in the corresponding Riemann sheet.
The overlapping curves represent a complex conjugate pair, for which the real part of the energy is exactly the same.
The vertical and horizontal gray lines represent the four van Hove singularities.
%
In the bottom figures~(e) and~(f), parts of the top four figures~(a)--(d) are shown with all Riemann sheets superimposed.
}
\label{FIG-realSolns}
\end{figure*}

We now analyze the detailed energy spectrum for our ladder model in the present $0 < t_\textrm{h}^\prime < t_\textrm{h}$ case.
In Fig.~\ref{FIG-realSolns} we present the real part of the twelve solutions of the dispersion equation~(\ref{adatomDisp}) as a function of the impurity energy $E_\textrm{d}$.
We have also plotted the line $\mathop{\mathrm{Re}} E=E_\textrm{d}$ that represents an unshifted energy for the adatom (the impurity energy if there was no interaction).  Hence, the deviation of each solution from this line represents the energy shift due to the interaction with the two-channel wire.  In these and the following figures we will always use $t_\textrm{h}=1$ as the unit of energy.

The behavior of the two purely real solutions P2 and R1 (Fig.~\ref{FIG-realSolns}(e)) are consistent with the energy spectrum previously pointed out~\cite{06TGP,05PTG,SG-Diss} for the persistent stable states mentioned above.  For values of $E_\textrm{d} \ll - t_\textrm{h} -t_\textrm{h}^\prime$ far below the lowest band edge, these two solutions are shifted downwards slightly
from the line $\mathop{\mathrm{Re}} E= E_\textrm{d}$.
(The shift for both solutions can be shown to be proportional to $g^2$, though P2 always has the slightly larger shift.)
For values $E_\textrm{d} \gg - t_\textrm{h} -t_\textrm{h}^\prime$ anywhere above the lowest band edge we find these solutions are shifted downwards instead from the lowermost band edge $\mathop{\mathrm{Re}} E= - t_\textrm{h} -t_\textrm{h}^\prime$, consistent with the previously reported behavior for the persistent stable state.  Below, we will show that in this case the shift is proportional to $g^4$.

\begin{figure*}
\hspace*{0.05\textwidth}
\includegraphics[width=0.4\textwidth]{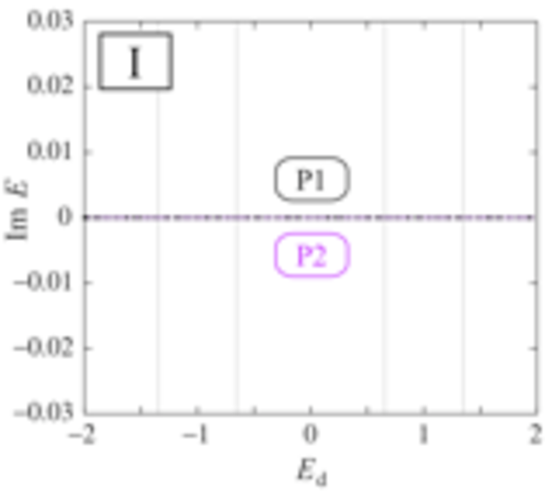}
\hfill
\includegraphics[width=0.4\textwidth]{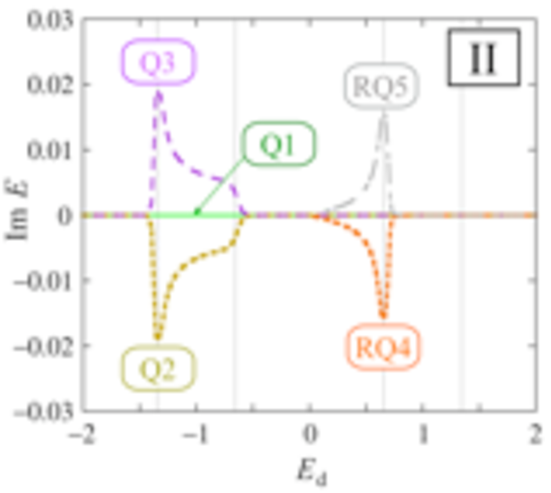}
\hspace*{0.05\textwidth}
\\
\vspace*{-\baselineskip}
\hspace*{0.08\textwidth}(a)\hspace*{0.465\textwidth}(b)\hspace*{0.4\textwidth}
\\
\vspace*{2\baselineskip}
\hspace*{0.05\textwidth}
\includegraphics[width=0.4\textwidth]{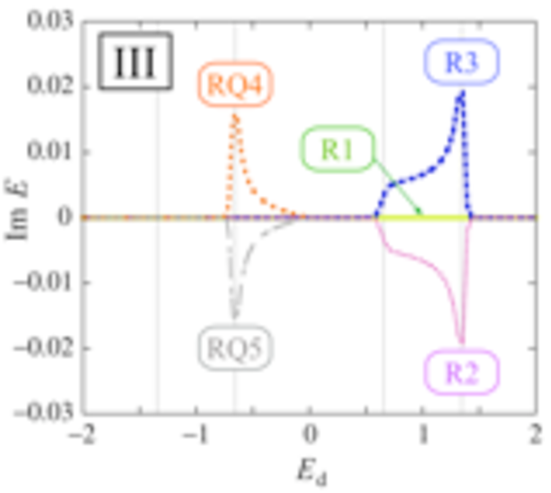}
\hfill
\includegraphics[width=0.4\textwidth]{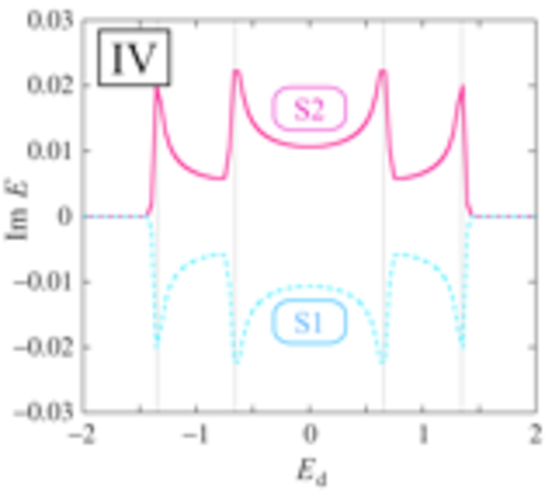}
\hspace*{0.05\textwidth}
\\
\vspace*{-\baselineskip}
\hspace*{0.08\textwidth}(c)\hspace*{0.465\textwidth}(d)\hspace*{0.4\textwidth}
\vspace*{\baselineskip}
\caption{Imaginary part of the energy for the eight complex solutions of the dispersion equation as a function of $E_\textrm{d}$ for the values $t^\prime_\textrm{h} = 0.345 t_\textrm{h}$ and $g = 0.1 t_\textrm{h}$.
The unit of the energy is $t_\textrm{h}=1.$ 
The vertical gray lines represent the four van Hove singularities.
}
\label{FIG-imagSolns}
\end{figure*}

\begin{figure}
\includegraphics[width=0.4\textwidth]{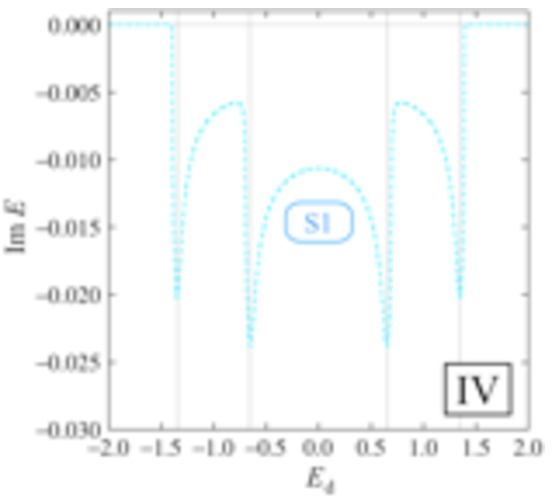}
\caption{Imaginary part of the energy (decay rate) for solution S1 (Sheet~IV) as a function of $E_\textrm{d}$ for the values $t^\prime_\textrm{h} = 0.345 t_\textrm{h}$ and $g = 0.1 t_\textrm{h}$.
The unit of the energy is $t_\textrm{h}=1.$
The vertical gray lines represent the four van Hove singularities, just as in Fig.~\ref{FIG-imagSolns}.
The decay rate for S1 is amplified (such that $\mathop{\mathrm{Im}}E \sim g^{4/3}$) in the vicinity of each of the band edges due to these singularities.
}
\label{FIG-imS1}
\end{figure}

For values $E_\textrm{d} \ll - t_\textrm{h} -t_\textrm{h}^\prime$ we find that the solutions S1 and S2 (lower left-hand corner of Fig.~\ref{FIG-realSolns}(d)) are also purely real.  These states may be called anti-bound states or virtual states~\cite{Humblet61}.  
As we increase the value of $E_{\textrm{d}}$ such that $E_\textrm{d}  \lesssim - t_\textrm{h} -t_\textrm{h}^\prime$ we find that these solutions merge abruptly to form a complex conjugate pair.  This is similar to the behavior of the complex solutions in the single channel model~\cite{05PTG}.  The imaginary part of these solutions can be seen in Fig.~\ref{FIG-imagSolns} (graphed with all the complex solutions), and in detail in Fig.~\ref{FIG-imS1}.
Notice in these figures that the decay rate is amplified for S1 and S2 in the vicinity of each of the four band edges.  This amplification is a result of the breakdown of Fermi's golden rule in the vicinity of the van Hove singularity at each of the band edges.  Ordinarily, the golden rule predicts that to lowest order the decay rate will be proportional to square of the coupling constant $g^2$; however, the van Hove singularity in the context of a one-dimensional system results in a decay rate with a non-analytic dependence on the coupling constant, such that to lowest order the decay rate is proportional to $g^{4/3}$, as previously reported~\cite{05PTG,06TGP}.

\subsection{Van Hove singularity and the origin of the quasi-bound states in continuum}

We will now consider the detailed behavior and the origin of the QBIC effect.  The decaying solutions Q2, R2 and RQ4 (or RQ5) (and their conjugate ``growth'' partners) each display this effect in different regions of the energy spectrum.  Here we will focus on the solution Q2 as our example in order to demonstrate the properties of the QBIC states.  Taking symmetry into account (see Fig.~\ref{FIG-imagSolns}), the solution R2 behaves in a manner almost analogous to Q2.
Meanwhile a detailed analysis of the integration contour for the survival probability of the excited state reveals that the solutions RQ4 and RQ5 may be of less significance in terms of the QBIC effect as it does not contribute a pole (exponential decay) in the most natural integration contour.
However, it is possible that this solution will play a more significant role in terms of the non-Markovian decay.

\begin{figure*}
\hspace*{0.05\textwidth}
\includegraphics[width=0.4\textwidth]{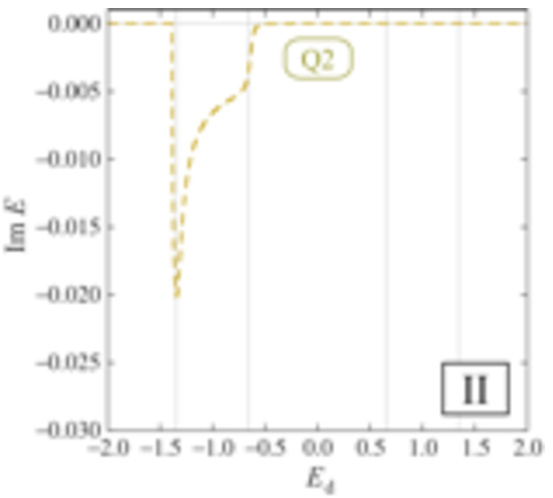}
\hfill
\includegraphics[width=0.4\textwidth]{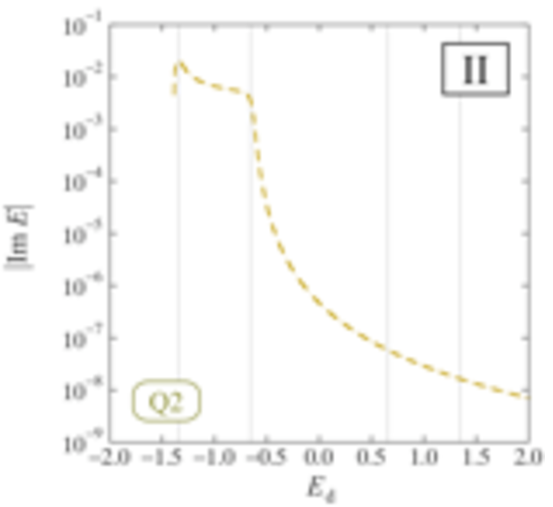}
\hspace*{0.05\textwidth}
\\
\vspace*{-\baselineskip}
\hspace*{0.08\textwidth}(a)\hspace*{0.465\textwidth}(b)\hspace*{0.4\textwidth}
\vspace*{\baselineskip}
\caption{Imaginary part of the energy (decay rate) for solution Q2 (Sheet~II) as a function of $E_\textrm{d}$ for the values $t^\prime_\textrm{h} = 0.345 t_{\textrm{h}}$ and $g = 0.1 t_{\textrm{h}}$.
(a) A linear plot on the and (b) a semi-logarithmic plot.
The unit of the energy is $t_\textrm{h}=1$.
The vertical gray lines represent the four van Hove singularities, just as in Fig.~\ref{FIG-imagSolns}.
For the solution Q2, the decay rate is amplified (such that $\mathop{\mathrm{Im}}E \sim g^{4/3}$) in the vicinity of the lowest outer band edge $- t_\textrm{h} -t_\textrm{h}^\prime$.
For values of $E_{\textrm d} \gg t_\textrm{h}  -t_\textrm{h}^\prime$, however, the decay rate becomes small (with $\mathop{\mathrm{Im}}E \sim g^6$) as the solution behaves as a QBIC state.
}
\label{FIG-imQ2}
\end{figure*}

Focusing on the solution Q2,  we see in Figs.~\ref{FIG-realSolns} and~\ref{FIG-imagSolns} that the solution is complex for values of the impurity energy $E_\textrm{d}$ just below the lower inner band edge $t_\textrm{h}-t_\textrm{h}^\prime$.  However, in Fig.~\ref{FIG-realSolns}  we see that as we increase the value of $E_\textrm{d}$, the real part of Q2 approaches the inner band edge $t_\textrm{h}-t_\textrm{h}^\prime$ in a manner similar to the persistent stable state P2 discussed above.  Meanwhile, the imaginary component associated with this solution does not vanish near the inner band edge.  Instead, the decay rate lies near the value  zero as can be seen in Fig.~\ref{FIG-imagSolns} and in greater detail in Fig.~\ref{FIG-imQ2}.
This is the QBIC state introduced in our previous Letter~\cite{QBIC-Letter}.  As can be seen from the figures, the real part of the energy is embedded in the lower energy band similar to the bound states in continuum proposed by von Neumann and Wigner, while the decay rate is non-zero but remarkably small.

We will now show that the QBIC effect is a direct result of two competing effects resulting from the embedding of the van Hove singularity at the edge of one conduction band in the continuum of the other band.  The first effect is the the tendency of the embedded singularity to create a persistent stable state; in other words, there would be an ordinary persistent stable state if it were not for the second energy band.
The second effect is the tendency of the embedding conduction band to destabilize an otherwise stable state.  In order to make this point explicit, we will obtain an analytic expansion for the energy eigenvalue of the persistent stable state in a single channel model (see Fig.~\ref{FIG-stableState}) and compare this term-by-term with a similar expansion for the QBIC state Q2 in the present case 
(see Fig.~\ref{FIG-state}).  This follows from our discussion in the previous Letter~\cite{QBIC-Letter}.

\subsubsection{Analytic approximation for the eigenenergy for the persistent stable state in single channel model}

We may write the Hamiltonian $\hat{ \mathcal H}_-$ for a single channel quantum wire~\cite{footnote1} with energy shifted by $t^\prime_\textrm{h}$ as
\begin{eqnarray}
\hat{ \mathcal H}_- = 
& & -\frac{t_\textrm{h}}{2} \sum_{x} \left(  |x+1 \rangle\langle x |+ |x \rangle\langle x+1 | \right) + t^\prime_\textrm{h}
 \nonumber \\
& & + \frac{g}{ \sqrt{2} } \left( | \textrm d \rangle\langle 0, 1 |+ |0, 1 \rangle\langle \textrm d |  \right)  
+ E_{\textrm d} | \textrm d \rangle\langle \textrm{d}| .   \label{decoupled.ham}
\end{eqnarray}
We have chosen the energy offset $t^\prime_\textrm{h}$ here such that the single energy band for this Hamiltonian mimics that of the upper energy band ($- t_\textrm{h} \cos K_{-} + t^{\prime}_\textrm{h}$) in the two-channel model.  This single channel will also have the same density of states function $\rho_-$ from Eq.~(\ref{DOS}).  Then the exact form of the discrete dispersion equation for an electron in the adatom in the single channel case~\cite{06TGP,Mahan,SG-Diss,Hatano07} is given by
\begin{equation}
z - E_{\rm d} - \frac{g^2}{2} \left[ \frac{1}{\sqrt{(z - t^{\prime}_\textrm{h})^2 - t_\textrm{h}^2}} \right] = 0.
\label{singleChanDisp}
\end{equation}
This is equivalent to a quartic dispersion polynomial after squaring.
Note the presence of the singularities in the third term at $z = \pm t_\textrm{h} + t^\prime_\textrm{h}$; these are a result of the singularities in the density of states function $\rho_-$ just as in the two-channel case.

\begin{figure}
\includegraphics[width=0.45\textwidth]{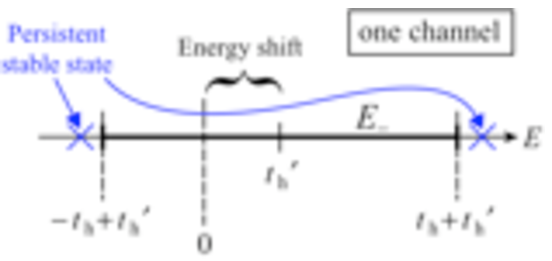}
\caption{Diagrammatic representation of the upper energy band $E_-$ associated with the (shifted) single channel Hamiltonian $\hat{ \mathcal H}_-$ and the purely real energy associated with the persistent stable state $E_{\textrm{ps}}$.}
\label{FIG-stableState}
\end{figure}

Now we will obtain an approximate form for the energy of the persistent stable state as a solution $E_{\textrm{ps}}$ of the single channel discrete dispersion equation~(\ref{singleChanDisp}).  This approximation will hold under the assumption that the impurity energy is much larger than that of the lower inner energy band, that is $E_{\rm d} \gg -t_\textrm{h} + t^\prime_\textrm{h}$.
The energy of this persistent stable state and its relation to the energy band $E_-$ are represented diagrammatically in Fig.~\ref{FIG-stableState}; the definition of $E_-$ is given in Eq.~(\ref{contDisp}).
Considering this observation, we write an expansion for $E_{\textrm{ps}}$ near the lower band edge $-t_\textrm{h} + t^\prime_\textrm{h}$ in powers of the coupling constant as
\begin{equation}
E_{\textrm{ps}} = (-t_\textrm{h} + t^\prime_\textrm{h}) + \chi_{\alpha} g^{\alpha} + \chi_{\gamma} g^{\gamma} + \dots ,
\label{psGenApprox}
\end{equation}
in which $0<\alpha<\gamma$,  $|\chi_i| \sim 1$ is independent of the coupling $g$ at every order, and the power of $g$ in each term is to be determined below.
  Notice that we have ``skipped'' the $\beta$ term in Eq.~(\ref{psGenApprox}); this is anticipation of the appearance of the decay rate in a similar approximation that we will perform for the QBIC state further on  (cf.\ Eq.~(\ref{Q2GenApprox})).

We can now use Eq.~(\ref{psGenApprox}) to write the dispersion equation~(\ref{singleChanDisp}) as
\begin{equation}
z - E_{\textrm d} 
= \frac{g^2}{2} \frac{1}{\sqrt{ -2 t_\textrm{h} \chi_{\alpha} g^{\alpha}  -2 t_\textrm{h} \chi_{\gamma} g^{\gamma} + O (g^{\alpha + \gamma}, g^{2 \alpha})  } }.
\label{psIntExp}
\end{equation}
We can expand this to obtain
\begin{eqnarray}  \label{psExpansion}
& & (-t_\textrm{h} + t^\prime_\textrm{h} - E_{\textrm d}) + \chi_{\alpha} g^{\alpha} 
\\
& \approx & \frac{g^{2 -\alpha / 2}}{2 \sqrt{- 2 t_\textrm{h} \chi_{\alpha} } }
        + O(g^{\gamma - \alpha + (2 - \alpha/2) }, g^{\alpha + (2 - \alpha/2)}  ).  \nonumber
\end{eqnarray}
The term in parentheses on the LHS and the first term on the RHS represent the two first-order terms; this can be shown to be the only consistent choice.  Since the term in parenthesis is zeroth order in $g$, equating these terms gives the condition $2 - \alpha/2 = 0$, or $\alpha = 4$, as well as
\begin{equation}
\chi_{\alpha} = - \frac{1}{8 t_\textrm{h} (t^{\prime}_\textrm{h} - t_\textrm{h} -  E_{\textrm d})^2}.
\label{chi_alpha}
\end{equation}
The second-order correction is then given by equating the second term on the LHS of Eq.~(\ref{psExpansion}) with the second term on the RHS; this gives $\gamma = 2 \alpha = 8$.  Making use of Eq.~(\ref{chi_alpha}), we can then write the expansion for the real energy for the persistent stable state Eq.~(\ref{psGenApprox}) as
\begin{equation}
E_{\rm  ps} = (-t_\textrm{h} + t^\prime_\textrm{h}) - \frac{1}{8 t_\textrm{h} (t^{\prime}_\textrm{h} - t_\textrm{h} -  E_{\textrm d})^2} g^4  + O(g^8).
\label{psApprox}
\end{equation}
We see that the energy shift from the band edge $-t_\textrm{h} + t^\prime_\textrm{h}$ is of order $g^4$.  Notice that it was the cancellation of the band edge term $-t_\textrm{h} + t^\prime_\textrm{h}$ when we plugged the expansion~(\ref{psGenApprox}) into Eq.~(\ref{singleChanDisp}) that resulted in the $g^{2 - \alpha /2}$ term in Eq.~(\ref{psExpansion}); hence the persistent stable state is a direct result of the divergent van Hove singularity at the band edge.  Also, note that $E_{\textrm{ps}}$ is purely real, including higher orders.

\subsubsection{Analytic approximation for the QBIC state Q2 in the two-channel model}

\begin{figure}
\includegraphics[width=0.45\textwidth]{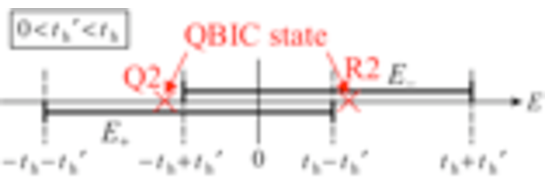}
\caption{Diagrammatic representation of the two energy bands $E_{\pm}$ associated with the full two-channel Hamiltonian $\hat{ \mathcal H}$ of Eq.~(\ref{ladder.ham}) and the real part of the energy associated with the quasi-bound state in continuum $E_{\textrm{Q2}}$ that lies below the lower inner band edge at $-t_\textrm{h} + t^\prime_\textrm{h}$.}
\label{FIG-state}
\end{figure}

\begin{figure}
\includegraphics[width=0.3\textwidth]{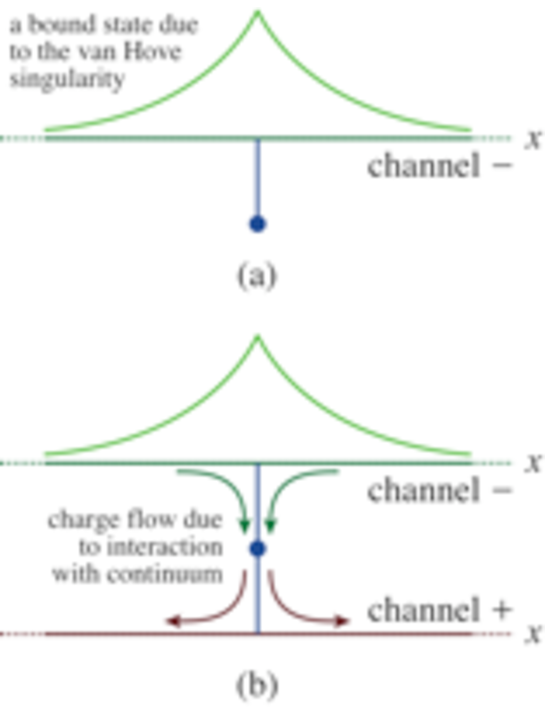}
\caption{
(a) A schematic view of the strongly bound state (due to the van Hove singularity) of a one-channel system with the eigenvalue just below the lower band edge.
(b) Some of the bound particles leak into the attached channel in the two-channel model.
}
\label{FIG-WF+-diag}
\end{figure}

Now we will obtain a similar expansion for the energy of the QBIC state Q2 in the case of the two-channel model that is the main subject of this paper.  We again present a diagrammatic representation of the energy of this state and its relation to the two energy bands in Fig.~\ref{FIG-state}.
As before, if we assume that $E_\textrm{d} \gg -t_\textrm{h} + t^\prime_\textrm{h}$, then the real part of the energy of this state will be shifted such that it lies slightly below the lower edge of the upper energy band at $-t_\textrm{h} + t^\prime_\textrm{h}$ just as in the case of the persistent stable state.
However, in this case the band edge (and therefore the energy of the state Q2 as well) is embedded in the continuum of the lower energy band.  It is well known that the continuum will have a de-stabilizing effect on a discrete state that is embedded within it.  In this case the embedding will result in a slight de-stabilization of the otherwise stable state.
For a further illustration of the relationship between the persistent stable state and the QBIC state, refer to Fig.~\ref{FIG-WF+-diag}(a,b).  

As before, we write the expansion for the energy of the state Q2 as 
\begin{equation}
E_{\rm Q2} = (-t_\textrm{h} + t^\prime_\textrm{h}) + \chi_{\alpha} g^{\alpha} + \chi_{\beta} g^{\beta} +  \chi_{\gamma} g^{\gamma} + \dots
\label{Q2GenApprox}
\end{equation}
where $0<\alpha<\beta<\gamma$.
Here we have included the $\beta$ term; below we will see that this term results in the small decay rate giving the QBIC effect.  The zeroth-order term $-t_\textrm{h} + t^\prime_\textrm{h}$ places the solution well inside the continuum of the lower energy band.  Looking to the discrete dispersion relation~(\ref{adatomDisp}), we note that it is the second term in the square brackets that is associated with the upper energy band $E_-$ including the embedded band edge at $-t_\textrm{h} + t^\prime_\textrm{h}$; the first term is associated with the embedding lower energy band $E_+$~\cite{footnote2}.

Putting the expansion~(\ref{Q2GenApprox}) into Eq.~(\ref{adatomDisp}), we obtain
\begin{eqnarray}
\lefteqn{
(-t_\textrm{h} + t^\prime_\textrm{h} - E_{\textrm d}) + \chi_{\alpha} g^{\alpha} 
}
\nonumber\\
 & \approx & \left( \frac{g^{2 -\alpha / 2}}{2 \sqrt{- 2 t_\textrm{h} \chi_{\alpha} } }
 	- \frac{\chi_{\beta} g^{\beta - \alpha + (2 - \alpha/2)} }{ 4 \sqrt{-2 t_\textrm{h} \chi_\alpha^3} }
	 \right.
\nonumber\\
&&\left.\phantom{\frac{g^{(\beta)}}{\sqrt{t_\textrm{h}\chi_\alpha^3}}}        +  g^{2 - \alpha/2} O(g^{\gamma - \alpha}, g^{\alpha}, g^{2(\beta - \alpha)}  )   
\right) \nonumber \\
& & + \left( \frac{g^2}{2 \sqrt{ ( 2 t^{\prime}_\textrm{h} - t_\textrm{h})^2 - t_\textrm{h}^2} } + O(g^{\alpha + 2})
        \right).
\label{Q2Expansion}
\end{eqnarray}
Again we note that in this expression the terms in the first set of parentheses on the RHS are associated with the embedded singularity at $-t_\textrm{h} + t^\prime_\textrm{h}$ and those in the second set of parentheses are associated with the embedding energy band.  We will see that equating one term from each results directly in the QBIC effect.
The first-order correction is obtained exactly as in the case of the persistent stable state above; by equating the term $-t_\textrm{h} + t^\prime_\textrm{h} - E_{\textrm d}$ on the LHS with the first term in the first set of parentheses we obtain the condition $\alpha = 4$ and Eq.~(\ref{chi_alpha}) as before.
The second-order condition is then obtained by equating the second term in the first set of parentheses on the RHS of Eq.~(\ref{Q2Expansion}) with the first term in the second set of parentheses on the RHS.  This gives the condition on $\beta$ as $\beta - \alpha = 2$, or $\beta = 6$, while we also obtain
\begin{equation}
\chi_{\beta} = \frac{\pm i}{ 16 t_\textrm{h}  (t^\prime_\textrm{h} -t_\textrm{h} - E_{\textrm d})^3 
\sqrt{ t^{\prime}_\textrm{h} (t_\textrm{h} -t^{\prime}_\textrm{h} )  } }.
\label{chi_beta}
\end{equation}
This correction is purely imaginary as $t_\textrm{h} -t^{\prime}_\textrm{h}  > 0$ in the case of overlapping bands.  This is the QBIC effect, appearing at the level of a second-order perturbation calculation.  It is a direct result of the interaction of the discrete state with the two overlapping energy bands, as we have just shown.  The remaining order terms in Eq.~(\ref{Q2Expansion}) yield the consistent result that $\gamma = 8$ as in the case of the persistent stable state before.

The expansion~(\ref{Q2GenApprox}) for the state Q2 can now be written as
\begin{eqnarray} \label{Q2Approx}
E_{\textrm{Q2}} & = & (-t_\textrm{h} + t^\prime_\textrm{h}) - \frac{1}{8 t_\textrm{h} (t^{\prime}_\textrm{h} - t_\textrm{h} -  E_{\textrm d})^2} g^4  \\
& & \pm i \frac{1}{ 16 t_\textrm{h}  (t^\prime_\textrm{h} -t_\textrm{h} - E_{\textrm d})^3 
\sqrt{ t^{\prime}_\textrm{h} (t_\textrm{h} -t^{\prime}_\textrm{h} )  } } g^6  \nonumber
+ O(g^8).
\end{eqnarray}
Comparison with Eq.~(\ref{psApprox}) above emphasizes that this state behaves essentially like the persistent stable state that results from the van Hove singularity; only in this case the energy shift puts this state in the continuum of the lower energy band with a small decay rate at order $g^6$.

For a numerical comparison, we can plug in the numbers $t^{\prime}_\textrm{h} = 0.345 t_\textrm{h}$, $g=0.1 t_\textrm{h}$, and $E_{\textrm d} = 0.3 t_\textrm{h}$ from Table~\ref{table.pole}.  Plugging these numbers into Eq.~(\ref{psApprox}) for the persistent stable state in the single channel case gives $E_{\textrm{ps}} \approx -0.655013701 t_\textrm{h}$ (this value includes the $g^8$ term, not explicitly given in Eq.~(\ref{psApprox}) above).  Plugging the same values into Eq.~(\ref{Q2Approx}) gives the value $E_{\textrm{Q2}} \approx -0.655013704 t_\textrm{h} - i (1.5095 \times 10^{-7}) t_\textrm{h}$ in agreement with the numerically obtained value reported in Table~\ref{table.pole}.

Finally, note that the expression for $E_{\textrm{Q2}}$ given above diverges in the case $t^{\prime}_\textrm{h} = t_\textrm{h}$. This is an indication that Eq.~(\ref{Q2Approx}) breaks down as $t^{\prime}_\textrm{h}$ approaches $t_\textrm{h}$.  We will find a new expression to replace Eq.~(\ref{Q2Approx}) in the special case $t^{\prime}_\textrm{h} = t_\textrm{h}$ in Sec.~IV.

\subsection{Wave function analysis and numerical simulations of time evolution}

In this subsection we will look more closely at the wave function for the QBIC state Q2.  In particular, we will show that the wave function for Q2 appears to be localized near $x=0$, although it actually behaves as a decaying state with an exponential divergence for large $x$.  
We will also verify that the ``$-$"~channel
 provides the dominant contribution to this wave function in the vicinity of the origin, as would be expected since it is the singularity associated with this channel that results in the nearly localized behavior of this state.
(Note that we have added quotation marks on the minus sign above to avoid any confusion in the notation. Hereafter, we will drop the quotation marks and simply write this as: $-$~channel.)
  Finally, we will also compare the time evolution of the ordinary decaying state S1 with that of Q2 in order to demonstrate that the QBIC decays on a much more gradual time scale than an ordinary decaying state.

\subsubsection{Wave function for QBIC state Q2 in the $y=1,2$ basis}

\begin{figure}
\begin{center}
\includegraphics[width=0.37\textwidth,clip]{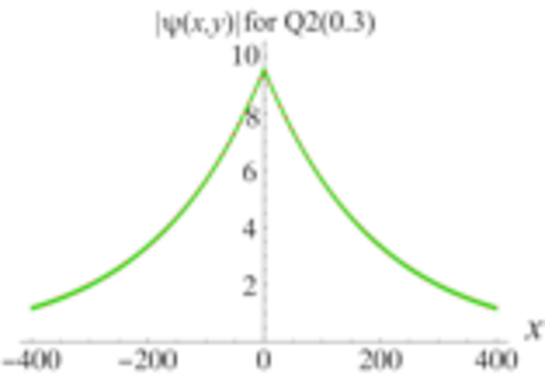}

(a)
\vspace*{\baselineskip}

\includegraphics[width=0.45\textwidth,clip]{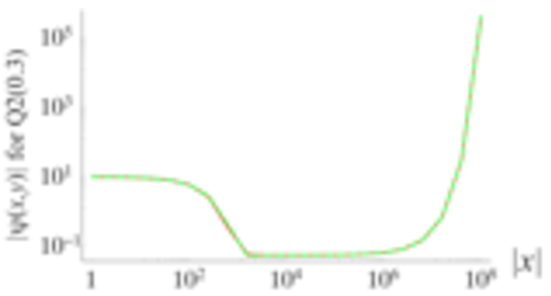}

(b)
\end{center}
\caption{(a) The wave function modulus $|\psi(x,y)|$ of the state Q2 around the origin (linear scale). (b) The same but away from the origin on the logarithmic scale.
The (overlapping) plots for $y=1$ (the upper leg) and $y=2$ (the lower leg) are almost indiscernible.
The parameters are set to $t'_\mathrm{h}=0.345t_\mathrm{h}$, $g=0.1t_\mathrm{h}$ and $E_\mathrm{d}=0.3t_\mathrm{h}$.
The wave function is normalized such that $\psi_\mathrm{d}=1$.}
\label{FIG-WF}
\end{figure}

In Fig.~\ref{FIG-WF}(a,b) we plot  a numerical result of
 the wave function from Eq.~(\ref{eq.40}) in the non-diagonalized channels $y=1,2$ for the state Q2.
Looking at Fig.~\ref{FIG-WF}(a) (linear scale) we see that the wave function for Q2 appears to be localized for values of $x$ near the origin (where the adatom is attached to the wire system).  However, in Fig.~\ref{FIG-WF}(b) (logarithmic scale) we see that the wave function indeed behaves as that for a decaying state with an exponential divergence in space far away from the origin.  Also notice that the contributions to the wave function from the two original channels $y= 1,2$ are almost exactly equal in either plot (the two graphs are overlapping).

\subsubsection{Wave function for QBIC state Q2 in the $\sigma=+,-$ basis}

\begin{figure}
\includegraphics[width=0.38\textwidth]{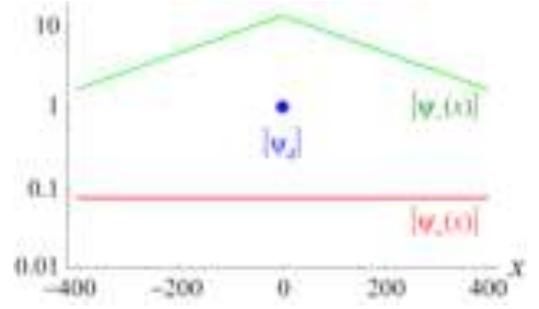}
\caption{The eigenfunction of the state Q2 for $t'_\mathrm{h}=0.345t_\mathrm{h}$, $g=0.1t_\mathrm{h}$ and $E_\mathrm{d}=0.3t_\mathrm{h}$ on a logarithmic scale as a function of position.
The amplitude modulus of the $-$~channel, $|\psi_-(x)|$, that of the $+$~channel, $|\psi_+(x)|$, and that of the dot, $|\psi_\mathrm{d}|$, are indicated.
The wave function is normalized such that $\psi_\mathrm{d}=1$.}
\label{FIG-WF+-}
\end{figure}

In Fig.~\ref{FIG-WF+-} we plot separately  a numerical result of the wave function contribution to the state Q2 from the $+$~and $-$~channels (as exemplified in Eq.~(\ref{eq.15})) in the basis of the partially diagonalized Hamiltonian~(\ref{two-channel.ham}).  
We also plot the discrete wave function associated with the impurity state, each at initial time $t=0$ with the usual choice of parameters $t'_\mathrm{h}=0.345t_\mathrm{h}$, $g=0.1t_\mathrm{h}$ and $E_\mathrm{d}=0.3t_\mathrm{h}$.  Keeping in mind the numerical choice for the coupling constant $g=0.1t_\mathrm{h}$, we can see in this figure that at the origin the amplitude of the wave function $|\psi_-(x)|$ for state Q2 in the $-$~channel is of the order $g$ larger than that of the adatom wave function $|\psi_\mathrm{d}|$, and that in turn $|\psi_\mathrm{d}|$ is order $g$ larger than the amplitude of the wave function $|\psi_+(x)|$ in the $+$~channel.  This implies that  $|\psi_+(x)|$ is order $g^2$ smaller than  $|\psi_-(x)|$, owing to the fact that the singularity in the $-$~channel gives the localized behavior of the state near the origin.

We may analytically demonstrate the relative orders of the wave functions by making use of the results that we obtained for the QBIC state previously in Sec.~II-B.  First we may add or subtract the first two equations in~(\ref{eq.46}) to obtain
\begin{equation}
\frac{|\psi_\mathrm{d}|}{|\psi_\pm(0)|} = \frac{ \sqrt{2} t_\mathrm{h} }{g} |\sin K_{\pm}|,
\label{magPsi}
\end{equation}
and the \emph{channel weight function}
\begin{equation}
\theta(E) \equiv \frac{|\psi_+(0)|}{|\psi_-(0)|} = \frac{| \sin K_- | }{ | \sin K_+ |} = \frac{\sqrt{{t_\mathrm{h}}^2-(E-t'_\mathrm{h})^2}}{\sqrt{{t_\mathrm{h}}^2-(E+t'_\mathrm{h})^2}},
\label{chanWeightFunc}
\end{equation}
in which we have made use of Eq.~(\ref{eq.14}) to write $\psi_\pm(0) = A_{\pm}$ at the origin $x=0$.  
These relations hold for any of the twelve eigenstates of the ladder system.

We here use the relations for the QBIC eigenvalue $E=E_\textrm{Q2}$ and the corresponding wave numbers $K_\pm$.
We can now use the expansion~(\ref{Q2Approx}) with the continuous dispersion equations~(\ref{contDisp}) to obtain
\begin{equation}
\sin K_+ = 2 \sqrt{  \frac{t^{\prime}_\mathrm{h}}{t_\mathrm{h}} \left(1 -  \frac{t^{\prime}_\mathrm{h}}{t_\mathrm{h}} \right) } + O(g^4)
\label{q2sinK_+}
\end{equation}
and
\begin{equation}
\sin K_- = i {1 \over 2 t_\mathrm{h} (t^{\prime}_\mathrm{h} - t_\mathrm{h} - E_\mathrm{d}) }g^2 + O(g^4).
\label{q2Sin_-}
\end{equation}
Applying these in Eq.~(\ref{magPsi}) gives 
\begin{eqnarray}
\frac{|\psi_\mathrm{d}|}{|\psi_\pm(0)|} \sim g^{\mp1}
\end{eqnarray}
in agreement with our discussion of Fig.~\ref{FIG-WF+-} above.  
Finally, the channel weight function~(\ref{chanWeightFunc}) for the state \textrm{Q2} is given by
\begin{equation}
\theta(E_\mathrm{Q2})={ |\psi_+(0)| \over |\psi_-(0)|} = \frac{g^2{t_\mathrm{h}}^2}{4 \sqrt{t^{\prime}_\mathrm{h} (t_\mathrm{h} - t^{\prime}_\mathrm{h})} \left|- t_\mathrm{h} + t^{\prime}_\mathrm{h} - E_\mathrm{d}\right| },
\label{q2PsiOverPsi}
\end{equation}
such that the contribution of the $+$~channel is order $g^2$ smaller than that of the $-$~channel.

\subsubsection{Generic channel weight function}

We may generalize the preceding discussion to states other than \textrm{Q2}.
As an application of Eq.~(\ref{chanWeightFunc}), consider the decaying states S1 and Q2 in the vicinity of the \emph{outer} band edge at $- t_\mathrm{h} - t^{\prime}_\mathrm{h}$.  Here the state Q2 will not behave as a QBIC state with a small decay rate, but instead has an amplified decay rate of order $g^{4/3}$ in the vicinity of the singularity at the outer band edge (for example, see Fig.~\ref{FIG-imQ2}(a))~\cite{06TGP}.  The state S1 also has an amplified decay rate for this range $E_\mathrm{d} \sim - t_\mathrm{h} - t^{\prime}_\mathrm{h}$.  For either of these solutions, Eq.~(\ref{chanWeightFunc}) gives 
$\theta(E_{\rm S1,Q2}) \sim g^{-2/3}$
, demonstrating that in this case, it is the lower channel $+$~that provides the largest contribution to the wave function.  The reason for this is that it is the van Hove singularity at the outer band edge $- t_\mathrm{h} - t^{\prime}_\mathrm{h}$ (associated with the lower band edge $E_+$) that results in the amplification of the decay rate, while the upper channel $E_-$ plays little role in this effect in this region of the energy spectrum.





\subsubsection{Time evolution for QBIC state Q2 against ordinary decaying state S1}

\begin{figure*}
\includegraphics[width=0.23\textwidth]{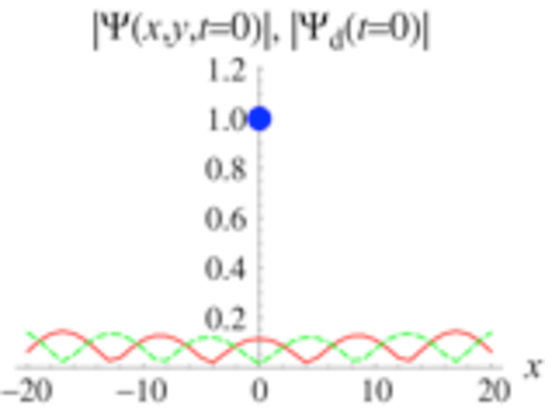}
\hfil
\includegraphics[width=0.23\textwidth]{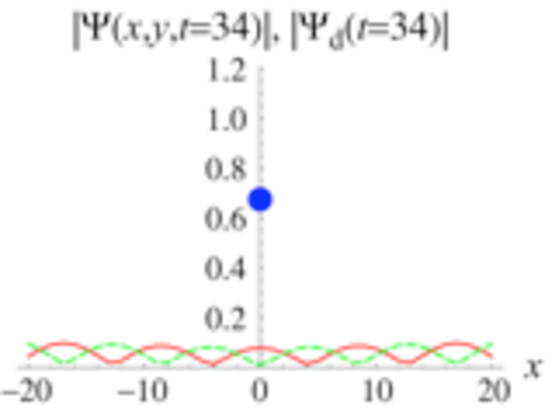}
\hfil
\includegraphics[width=0.23\textwidth]{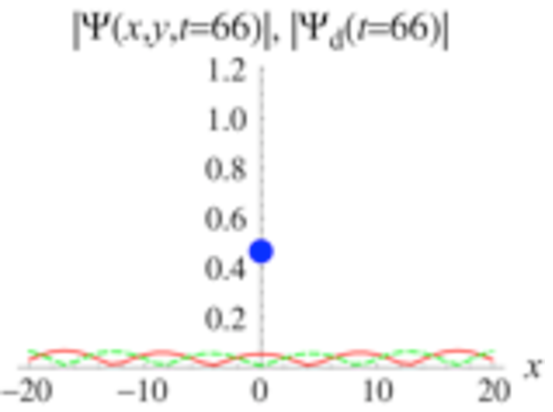}
\hfil
\includegraphics[width=0.23\textwidth]{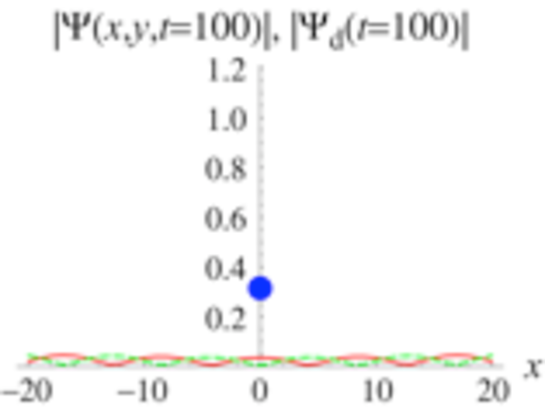}
\\
\hspace*{0.11\textwidth}(a)\hfill(b)\hfill(c)\hfill(d)\hspace*{0.12\textwidth}
\\
\begin{center}
Solution S1 for $E_\textrm{d}=0.3t_\textrm{h}$
\end{center}
\vspace{\baselineskip}
\includegraphics[width=0.23\textwidth]{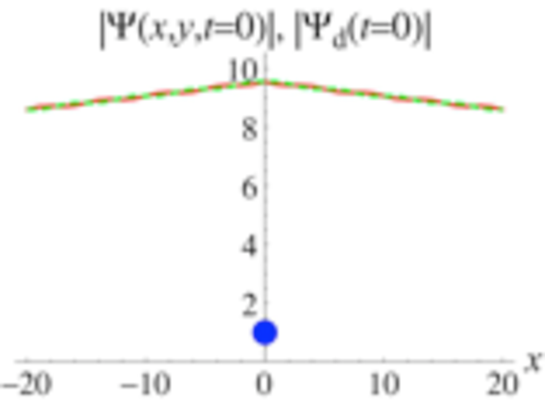}
\hfil
\includegraphics[width=0.23\textwidth]{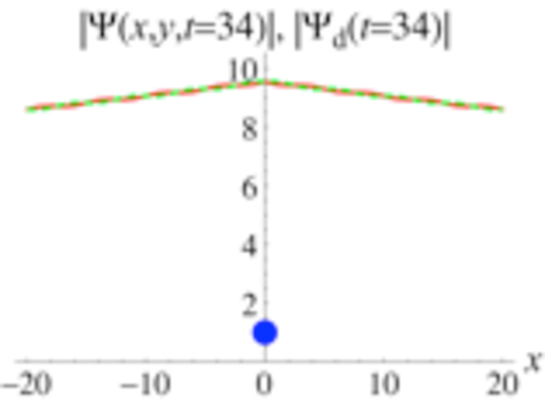}
\hfil
\includegraphics[width=0.23\textwidth]{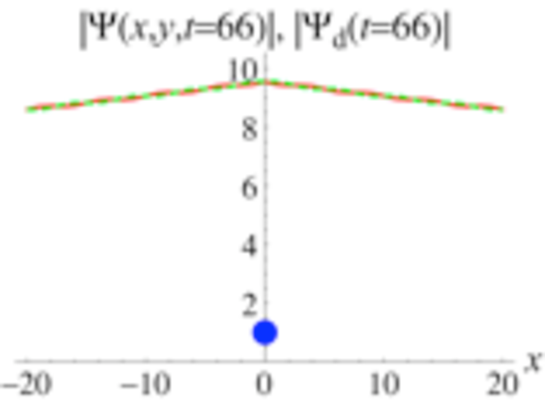}
\hfil
\includegraphics[width=0.23\textwidth]{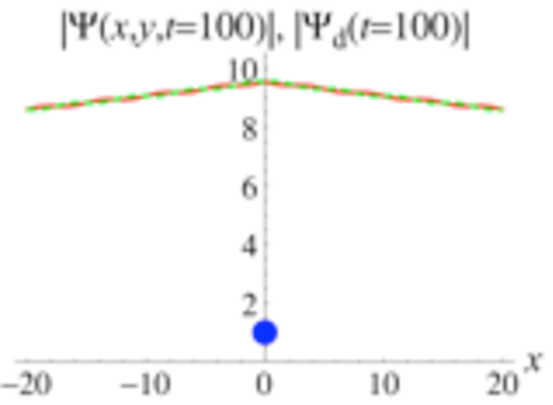}
\\
\hspace*{0.11\textwidth}(e)\hfill(f)\hfill(g)\hfill(h)\hspace*{0.12\textwidth}
\\
\begin{center}
Solution Q2 for $E_\textrm{d}=0.3t_\textrm{h}$
\end{center}
\caption{Time dependence of wave function of solutions S1~((a)--(d)) and Q2~((e)--(h)) 
for $t^\prime_\textrm{h} = 0.345 t_{\textrm{h}}, \ g = 0.1t_{\textrm{h}}$  and $E_\textrm{d}=0.3 t_{\textrm{h}}$.  
The solid (red) curves represent $\Psi(x,1,t)$, the broken (green) curves represent $\Psi(x,2,t)$ and the dots represent $\Psi_\mathrm{d}(t)$.
The wave functions are normalized such that $\Psi_\mathrm{d}(0)=1$.
The QBIC solution Q2 clearly decays on a much slower time scale than that of the ordinary decay state S1.}
\label{FIG-timeComp}
\end{figure*}

In Fig.~\ref{FIG-timeComp} we show the time evolution $\Psi(x,y,t)$ and $\Psi_\mathrm{d}(t)$ for the states S1 and Q2 for the ordinary choice of the parameters $t^\prime_\textrm{h} = 0.345 t_{\textrm{h}}, \ g = 0.1 t_{\textrm{h}}$  and $E_\textrm{d}=0.3 t_{\textrm{h}}$, under which the state Q2 will behave as a QBIC state.  We see that indeed on the time scale under which the ordinary state S1 decays almost completely, the state Q2 appears to behave as a localized state without a noticeable decay rate (for this time scale).  
The details of the numerical method by which we have obtained the time evolution simulations in these plots are presented in Appendix A.

\section{Energy spectrum analysis in two special cases}

In this section, we will briefly consider the energy spectrum analysis for two special cases.  In the first special case ($t^\prime_\textrm{h} = 0$) the two-channel model reduces to the single channel model when the chain-to-chain hopping parameter $t^\prime_\textrm{h}$ vanishes.  
Even though this case is trivial, it is instructive to see the relation of the two-channel model to the single channel model.  Then we will consider the case 
$t^\prime_\textrm{h} = t_\textrm{h}$. Here we will find that the 12th-order dispersion polynomial reduces to a 10th-order polynomial, while the QBIC decay rate is amplified such that it is proportional to $g^4$ to first order; both embedded singularities play a role in this modified QBIC effect.

\subsection{Energy spectrum for $t^\prime_\textrm{h} = 0$ case}

Here we will comment on the case $t^\prime_\textrm{h} = 0$, in which the chain-to-chain hopping parameter vanishes.  Since the adatom is coupled to only one chain, it is to be expected that this system should reduce to the single chain model (along with an additional uncoupled chain).  Indeed, considering the discrete dispersion equation~(\ref{adatomDisp}) we see that for $t^\prime_\textrm{h} = 0$  the two terms on the RHS containing the square roots will agree.  Therefore, in the case where the sign of the two square roots agrees, these two terms will combine to give a new dispersion equation equivalent to that for the single-channel model (equal to Eq.~(\ref{singleChanDisp}) after setting $t^\prime_\textrm{h} = 0$ and replacing with the re-normalized coupling constant $g^{\prime} \rightarrow \sqrt{2}g$
).  The case in which the sign of these two terms agree corresponds to Sheets I and IV in the complex energy surface, which contain four solutions (two in each sheet).  These four solutions then behave precisely as the four solutions to the quartic dispersion polynomial in the original single-channel model.

Meanwhile, for the case in which the sign of the two square roots in Eq.~(\ref{adatomDisp}) are opposite, then these two terms cancel and the dispersion equation becomes trivial.  This case corresponds to Sheets II and III, which contain eight solutions.  Meanwhile, since the two branch cuts in this case overlap exactly, Sheets II and III become mathematically (and physically) inaccessible.  If one travels through the branch cut in Sheet~I, one will always appear in Sheet~IV (and vice versa).  Hence the energy surface is effectively two-sheeted and essentially equivalent to that in the single chain system.

\subsection{Energy spectrum for $t^\prime_\textrm{h} = t_\textrm{h} $ case}

\begin{figure}
\includegraphics[width=0.45\textwidth]{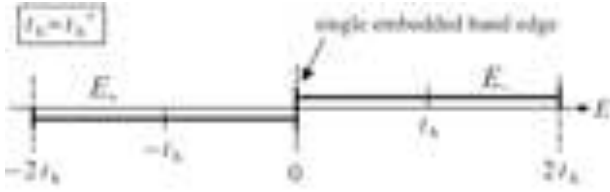}
\caption{Band structure for the $t_\textrm{h}^\prime = t_\textrm{h}$ case with a single ``unified'' embedded band edge in the center of the spectrum.}
\label{FIG-bandsOverlap}
\end{figure}

In the case $t^\prime_\textrm{h} = t_\textrm{h}$, the two inner band edges will overlap to form a single embedded band edge as indicated in Fig.~\ref{FIG-bandsOverlap}.  
Note that, in a certain sense, both band edges are still present.  In fact, we will find that in this case the van Hove singularity from one overlapping band edge will actually amplify the QBIC decay rate that results from the van Hove singularity of the other band edge, so that the decay rate will now be proportional to $g^4$.  This is a unique combination of two previous effects, both resulting from the van Hove singularity.

\subsubsection{Tenth order dispersion polynomial for the $t^\prime_\textrm{h} = t_\textrm{h}$ case}

If we set $t^\prime_\textrm{h} = t_\textrm{h}$ in Eq.~(\ref{adatomDisp}), then the discrete dispersion equation becomes
\begin{equation}
z - E_{\rm d} - \frac{g^2}{2 \sqrt{|z|} } \left[ \frac{1}{\sqrt{z + 2t_\textrm{h}}}  + \frac{1}{\sqrt{z - 2t_\textrm{h} }} \right] = 0.
\label{adatomDispOverlap}
\end{equation}
Note that here the position of a solution in the complex energy surface should be determined only by the two square roots $1 / \sqrt{z \pm 2t_\textrm{h}}$; the overall factor of $1/ \sqrt{|z|}$ plays no role in this determination.  However, this overall factor represents the presence of a van Hove singularity at $z = 0$ in both bands.

As before, we can square this equation twice to find an equivalent tenth-order dispersion polynomial; hence two solutions have vanished from the system in comparison to the general case.  Specifically, the two solutions RQ4 and RQ5 are no longer present.  Note that the QBIC solutions Q2 and R2 remain in this simplified system.  

As was done in Sec.~III~A and in Fig.~\ref{fig-K+-}, it is more convenient for the purpose of placement of each solution in the correct Riemann sheet to solve
\begin{eqnarray}
\lefteqn{
-t_\textrm{h} \cos K_+ -t_\textrm{h}
=-t_\textrm{h} \cos K_- +t_\textrm{h}
}
\nonumber
\\
& = &E_{\textrm d} +g^2
\left(
\frac{1}{2it_\textrm{h} \sin K_{+}} + 
\frac{1}{2it_\textrm{h} \sin K_{-}} 
\right). \label{eq.49-3}
\end{eqnarray}
with respect to $K_\pm$ than to solve the discrete dispersion equation~(\ref{adatomDispOverlap}) with respect to $z=E$.
The imaginary parts of the solutions of the above simultaneous equations give the correct Riemann sheet.
The eigenenergy of each solution is given by the continuous dispersion equation $E=-t_\textrm{h}\cos K_\pm \mp t_\textrm{h}$.

\begin{figure*}
\hspace*{0.05\textwidth}
\includegraphics[width=0.4\textwidth]{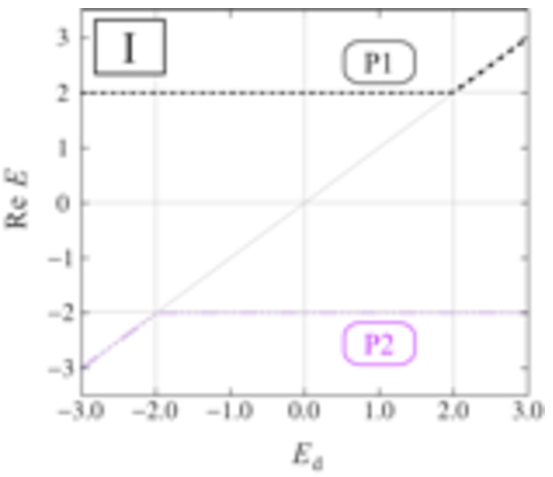}
\hfill
\includegraphics[width=0.4\textwidth]{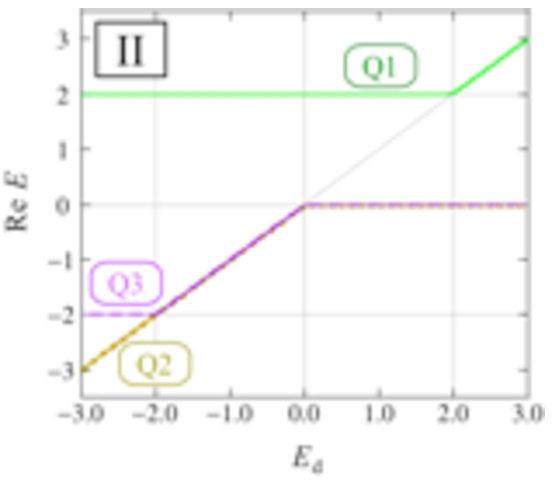}
\hspace*{0.05\textwidth}
\\
\vspace*{-\baselineskip}
\hspace*{0.08\textwidth}(a)\hspace*{0.465\textwidth}(b)\hspace*{0.4\textwidth}
\\
\vspace*{2\baselineskip}
\hspace*{0.05\textwidth}
\includegraphics[width=0.4\textwidth]{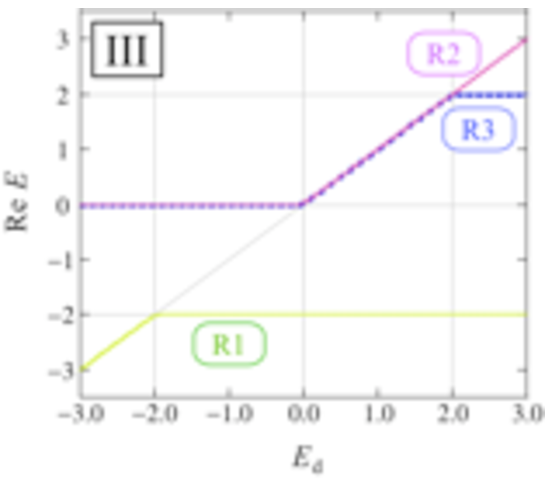}
\hfill
\includegraphics[width=0.4\textwidth]{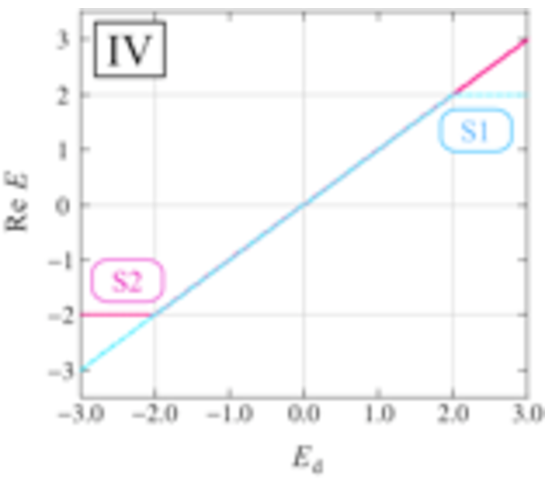}
\hspace*{0.05\textwidth}
\\
\vspace*{-\baselineskip}
\hspace*{0.08\textwidth}(c)\hspace*{0.465\textwidth}(d)\hspace*{0.4\textwidth}
\\
\vspace*{2\baselineskip}
\hspace*{0.05\textwidth}
\centering
\includegraphics[width=0.4\textwidth]{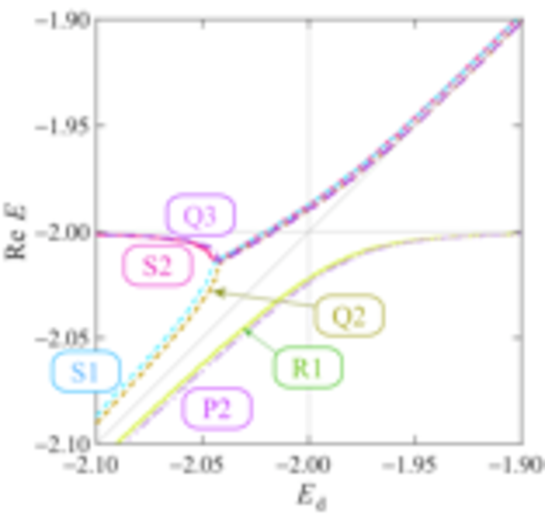}
\hspace*{0.05\textwidth}
\\
\vspace*{-\baselineskip}
\hspace*{0.34\textwidth}(e)\hspace*{0.65\textwidth}
\vspace*{\baselineskip}
\caption{Real part of the energy for the ten solutions of the simplified dispersion equation~(\ref{adatomDispOverlap}) as a function of $E_\textrm{d}$ for the special case  $t^\prime_\textrm{h} = t_\textrm{h}$ with the choice $g = 0.1 t_\textrm{h}$.
The unit of the energy is $t_\textrm{h}=1$.
In the top four figures~(a)--(d), each solution is plotted in the corresponding Riemann sheet.
The overlapping curves represent a complex conjugate pair, for which the real part of the energy is exactly the same.
The vertical and horizontal gray lines represent the three van Hove singularities.
In the bottom figure~(e), a part of the top four figures~(a)--(d) are shown with all Riemann sheets superimposed.
}
\label{FIG-realSolnsOverlap}
\end{figure*}

\begin{figure*}
\hspace*{0.05\textwidth}
\includegraphics[width=0.4\textwidth]{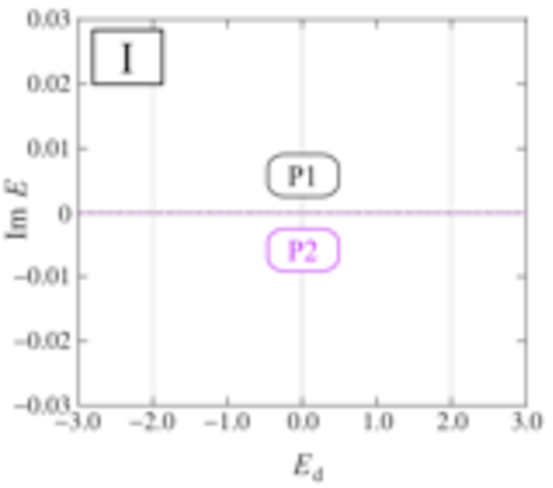}
\hfill
\includegraphics[width=0.4\textwidth]{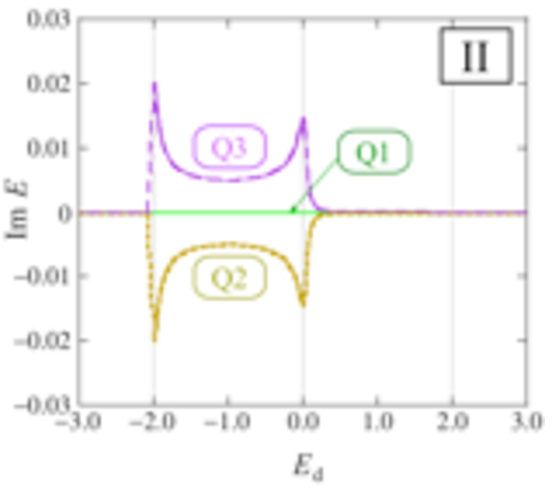}
\hspace*{0.05\textwidth}
\\
\vspace*{-\baselineskip}
\hspace*{0.08\textwidth}(a)\hspace*{0.465\textwidth}(b)\hspace*{0.4\textwidth}
\\
\vspace*{2\baselineskip}
\hspace*{0.05\textwidth}
\includegraphics[width=0.4\textwidth]{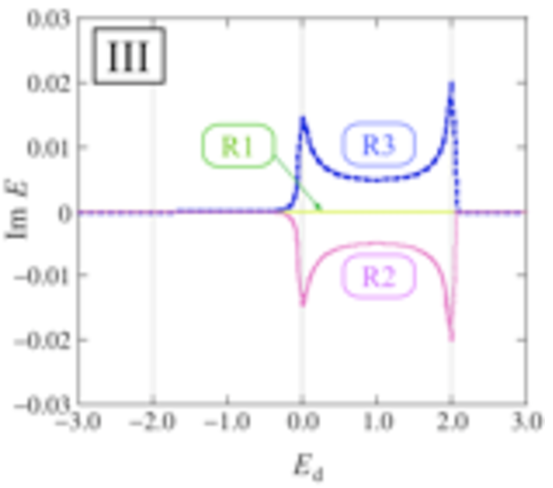}
\hfill
\includegraphics[width=0.4\textwidth]{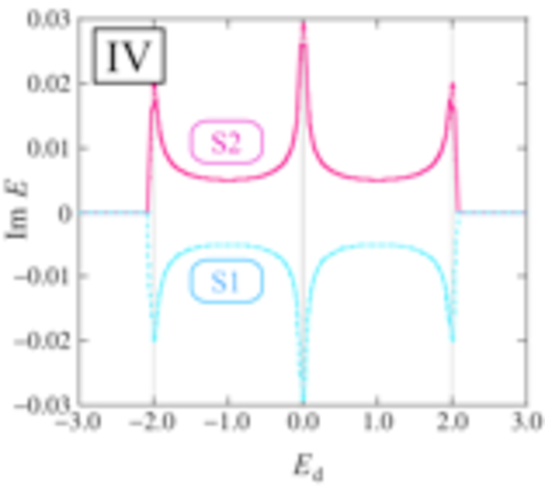}
\hspace*{0.05\textwidth}
\\
\vspace*{-\baselineskip}
\hspace*{0.08\textwidth}(a)\hspace*{0.465\textwidth}(b)\hspace*{0.4\textwidth}
\vspace*{\baselineskip}
\caption{Imaginary part of the energy for the six complex solutions of the simplified dispersion equation~(\ref{adatomDispOverlap}) as a function of $E_\textrm{d}$ for the special case $t^\prime_\textrm{h} = t_\textrm{h}$ with the choice $g = 0.1 t_\textrm{h}$.
The unit of the energy is $t_\textrm{h}=1.$
The vertical gray lines represent the three van Hove singularities.
}
\label{FIG-imagSolnsOverlap}
\end{figure*}

The dependence of the real and imaginary parts of all the solutions on $E_\textrm{d}$ can be seen in Figs.~\ref{FIG-realSolnsOverlap} and~\ref{FIG-imagSolnsOverlap}, respectively, for the value $g=0.1  t_\textrm{h}$ for the coupling.

\subsubsection{Modified QBIC effect}

As in the general case, the solutions Q2 (for $E_\textrm{d} \gtrsim t_\textrm{h}$) and R2 (for $E_\textrm{d} \lesssim -t_\textrm{h}$) are QBIC states.  Their real parts behave similar to the persistent stable states in that they lie near to the embedded band edge at $z = 0$ in Fig.~\ref{FIG-realSolnsOverlap}.  However, they have a small non-zero decay rate as can be seen in Fig.~\ref{FIG-imagSolnsOverlap}.  We can obtain an approximate form for the energy eigenvalue for these states in a manner similar to that we employed before in the general case.  
Figure~\ref{FIG-realSolnsOverlap} shows that the real part of the energy for Q2 is also small around the origin.  Hence, we assume that the expansion of $E_{\textrm{Q2}} $ begins with the order $g^\alpha$ with $\alpha > 0$ as $E_{\textrm{Q2}} = \chi_\alpha g^\alpha + \cdots$.
Applying this expansion in Eq.~(\ref{adatomDispOverlap}) 
and using a similar argument as that given in Sec.~III~B
gives
\begin{equation}
\sqrt{\chi_\alpha} = g^{2 - \alpha/2} \frac{1}{2 \sqrt{2 t_\textrm{h}} E_\textrm{d} } ( 1 - i ),  
\label{eq.43}
\end{equation}
in which the two bracketed terms (both of which are necessary for $E_{\rm Q2}$ to be imaginary) have been contributed respectively by the two bracketed terms on the LHS of~(\ref{adatomDispOverlap}), and the factor $1/ \sqrt{|z|}$ associated with the van Hove singularity in \emph{both} terms has resulted in the factor $g^{2 - \alpha/2}$.  We then obtain the familiar condition for the order of $\alpha$ as $\alpha = 4$ as well as the condition $\chi_\alpha = {-i}/({4 t_\textrm{h} E_\textrm{d}^2})$, from which we obtain
\begin{equation}
E_{\textrm{Q2}} = - i \frac{g^4}{4 t_\textrm{h} E_\textrm{d}^2} + O(g^8).
\label{Q2ApproxOverlap}
\end{equation}
We see that the state Q2 is still quasi-stable in comparison to the ordinary decay rate proportional to $g^2$.  However, the QBIC decay rate has been amplified, so that it is proportional to $g^4$ (instead of $g^6$ in the general case) due to the overlapping band edges.  In a sense, the embedded van Hove singularity at $z=0$ from the $E_-$ band has resulted in a QBIC state while the singularity from the $E_+$ band has amplified the usual decay rate of $g^6$ to $g^4$.

Indeed, the channel weight function~(\ref{chanWeightFunc}) reduces to
\begin{equation}
\theta(E) =
\sqrt{  E + 2 t_\textrm{h}  \over
  	  E  - 2 t_\textrm{h}}
\end{equation}
for $t^{\prime}_\textrm{h} = t_\textrm{h}$ and gives $\theta(E_{\rm  Q2}) \sim i$ for the specific case of the QBIC state.  Thus both channels contribute equally in this modified QBIC effect.


\section{Concluding remarks}

Before we conclude this paper let us give a brief comment on the order of the dispersion polynomial and number of solutions (and QBIC solutions) for an $n$-channel wire model.
In the single channel model~\cite{06TGP,SG-Diss,Hatano07} the quartic dispersion polynomial yielded four solutions.  If we consider the position of the real part of these solutions (see, for example, Fig.~2(a) in~\cite{06TGP}), then regardless of the value of $E_\textrm{d}$ there are always two solutions that lie near the value $E_\textrm{d}$ and one solution that lies near each of the two band edges; when $E_\textrm{d}$ lies well within the conduction band, both of the latter solutions are persistent stable states that co-exist with the decay solution.  We can conceptualize this system with the statement that we have $2_{E_\textrm{d}} + 2_\mathrm{BE} = 4$ solutions, where $2_{E_\textrm{d}}$ indicates that there are two solutions near $E_\textrm{d}$ and $2_\mathrm{BE}$ indicates that there are two more solutions near the two band edges of a channel.

In general, there are two band edges associated with each channel and hence the total number of the band edges is $N_\mathrm{BE} = 2n$.
In the present case we can see in Fig.~\ref{FIG-realSolns} that for the two-channel model there are always four solutions that lie near the energy $E_\textrm{d}$ ($4_{E_\textrm{d}}$), while there are two solutions near each of the four band edges ($2 \times 4_\mathrm{BE}$); hence  we have $4_{E_\textrm{d}} + 2 \times 4_\mathrm{BE} = 12$ solutions.  We can generalize this observation by saying that for an $n$-channel model there are $(2^n)_{E_\textrm{d}}$ solutions associated with the discrete energy $E_\textrm{d}$ and $2^{n-1} \times N_\mathrm{BE} = 2^n n$ solutions associated with the $N_\mathrm{BE}$ band edges.  Hence there are $N_{\textrm{solns}} = 2^n (1+n)$ total solutions for an $n$-channel quantum wire coupled with a single discrete state, given as the solutions to an $N_{\textrm{solns}}$-order dispersion polynomial.  As mentioned in Sec.~II~B, these solutions will live in a Riemann surface composed of $2n$ Riemann sheets.

Regarding the number of QBIC states for the $n$-channel model, this number will depend not only on the number of embedded band edges, but also on the value of $E_\textrm{d}$.  However, we can state that in general the maximum number of QBIC states for each model is equal to the number of band edges which could possibly be embedded in the continuum of another band.  This number is equal to $2^{n-1}$, although this equation is obviously not valid in the single-channel model for which no embedding is possible. 
We intend to present these statements in greater detail elsewhere.

We have demonstrated the existence of the QBIC state in the context of a two-channel quantum wire with an attached adatom impurity.  Ordinarily, an electron in the impurity with an energy deep inside of the conduction band of the wire would be expected to decay and travel along the length of the wire.  We would also expect that we could describe the decay rate using Fermi's golden rule.  However, due to the combined effect of two overlapping conduction bands (with van Hove singularities at the band edges), we have shown that a QBIC electron will remain stable inside the impurity for ordinary time scales.  In particular, we have demonstrated the connection between the QBIC state and the persistent stable state that results from the van Hove singularity at the band edge in the single channel model.  In the two-channel model, this persistent stable state is slightly de-stabilized by the presence of a second conduction band.

We have also shown in Eq.~(\ref{Q2Approx}) that the characteristic decay rate for the QBIC is generally on the order of $g^6$, although this may be modified under certain conditions, such as the case $t^{\prime}_\textrm{h} = t_\textrm{h}$ in the two-channel model (under which $\mathop{\mathrm{Im}} E\sim g^4$). 

While we have demonstrated the above specifically for the two-channel quantum wire, it is easy to show that this effect should occur in other one-dimensional models which have the characteristic square root divergence in the DOS given in Eq.~(\ref{DOS}).  This includes models such as an electromagnetic waveguide~\cite{05PTG} when we consider two overlapping field modes, as we mentioned above.  Hence, the origin of the QBIC effect is quite different than that of the BIC effect originally proposed by von Neumann and Wigner.  While we can associate each QBIC state with a divergent band edge singularity embedded in the continuum of another energy band, the BIC states are associated with zeros in the interaction potential that occur in the continuous energy spectrum for certain models with an oscillating potential.  Hence the QBIC is more closely associated with the DOS function while the BIC is more closely associated with the interaction potential.  It is also conceivable that the QBIC effect may appear in some two-dimensional systems (such as a two-dimensional tight-binding lattice) that have a characteristic logarithmic divergence in the DOS~\cite{53vanHove}.

As we remarked in our previous Letter~\cite{QBIC-Letter}, because our quasibound state has a small decay rate (imaginary component of the eigenenergy), it is not, strictly speaking, ``in continuum.''  However, this decay rate is extremely small, such that the QBIC state should behave as if it were a bound state with real part of the eigenenergy deeply embedded in the continuum even on relatively large time scales.  In this sense, the QBIC will essentially behave as the BIC under actual experimental conditions.  Meanwhile, the BIC is a true bound state with a purely real energy spectrum, but only under ideal conditions.  Since the BIC exists only at discrete points (with zero measure) in parameter space any noise in the system (such as thermal noise) in an experiment may actually lead to a small decay rate for the BIC.  On the contrary, since the QBIC exists for a wide range of parameter space (with non-zero measure), it is robust against noise. It may be easier to prepare the QBIC in the experiment.

It should be mentioned that
 there may be certain models with embedded singularities in which the QBIC effect due to the DOS singularities will be washed out as a result of the form of the interaction potential.  There is at least one example of a single channel model in which the interaction potential washes out the effects of the singularity and prevents the persistent stable state from forming~\cite{TGOP07}.

\begin{acknowledgments} 
The authors thank Professor Satoshi Tanaka  and Professor E.~C.~G.~Sudarshan for useful discussions.  S.~G.\ would like to thank the National Science Foundation and the Japan Society for the Promotion of Science for their support, as well as Professor Satoshi Tanaka for his hospitality during a stay in Japan.  This material is based upon work supported by the National Science Foundation under Grant No.~0611506.
The work is supported partly by the Murata Science Foundation as well as by the National Institutes of Natural Sciences undertaking Forming Bases for Interdisciplinary and International Research through Cooperation Across Fields of Study and Collaborative Research Program (No.~NIFS08KEIN0091) and Grants-in-Aid for Scientific Research (No.~17340115, No.~17540384, and No.~20340101)) from the Ministry of Education, Culture, Sports, Science and Technology.
N.~H.\ acknowledges  support by Core Research for Evolutional Science and Technology (CREST) of Japan Science and Technology Agency.
\end{acknowledgments}

\appendix

\section{Numerical method for time evolution simulation of wave functions}

In this appendix we describe our numerical method for obtaining the time evolution of the wave functions of the resonant states of the two-channel Hamiltonian.  We rely on the method which was proposed in previous work~\cite{Hatano07} to solve the time-dependent Schr\"odinger equation accurately  (despite truncation of the domain of $x$ in numerical calculations).
The time-dependent Schr\"odinger equation is given by
\begin{equation}
i \hbar \frac{\partial }{\partial  t} | \Psi (t) \rangle = \hat{\mathcal H} |\Psi (t) \rangle, \label{eq.61}
\end{equation}
and the initial condition is fixed as
\begin{equation}
|\Psi (0) \rangle = |\psi \rangle, \label{eq.61a}
\end{equation}
where $|\psi \rangle$ is the eigenstate of $\cal{H}$ in Eq.~(\ref{Sch-eqn}).
We define the time-dependent wave function as follows:
\begin{equation}
\Psi_{\textrm d} (t) \equiv   \langle \textrm{d}|\Psi (t) \rangle ,  \ \ 
{\Psi} (x,y;t) \equiv  \langle x,y |\Psi (t) \rangle,
\label{eq.62} 
\end{equation}
where $y =1, 2$. 
The vector form of the wave functions is given by
\begin{eqnarray}
\vec{\Psi}_y (x,t) \equiv
\begin{pmatrix}
{\Psi} (x,1;t)\\
{\Psi} (x,2;t)
\end{pmatrix}
= 
e^{-iEt}  \vec{\psi}_y (x) .
\label{eq.63}
\end{eqnarray}
We restrict the region $|x|\le L, (L\ge 1)$ to compute the time evolution of the wave function.
The steady wave function $\vec{\psi}(x)$ in Eq.~(\ref{eq.40}) has the following recursion property:
\begin{eqnarray}
\vec{\psi}_y (L+1) 
&=& U \vec{\psi}_{\sigma} (L+1) \nonumber \\
&=& U \left\{ A_{+} e^{iK_{+} (L+1)} 
				\begin{pmatrix}
					1 \\
					0
				\end{pmatrix} 
				+ A_{-} e^{iK_{-} (L+1) } 
				\begin{pmatrix}
					0 \\
					1
				\end{pmatrix}  
		\right\} \nonumber \\  
&=& U \left(
			\begin{array}{cc}		
				e^{i K_{+}}   &       0     \\
			       0          & e^{iK_{-}} 
			\end{array}
		\right) U^{-1}  U   \  \vec{\psi}_{\sigma} (L) \nonumber \\
&=&  V_{\textrm{eff}} \  \vec{\psi}_y (L) ,
\label{eq.64}
\end{eqnarray}
where we have used Eqs.~(\ref{eq.15}) and~(\ref{eq.40}). 
We have defined the effective potential $V_\textrm{eff}$ as 
\begin{eqnarray}
 V_{\textrm{eff}} 
&\equiv&  U \left(
			\begin{array}{cc}		
				e^{i K_{+}}   &       0     \\
			       0          & e^{iK_{-}} 
			\end{array}
		\right) U^{-1}    \nonumber \\
&=&  \frac{1}{2}
\left(
\begin{array}{cc}
e^{i K_{+}} + e^{iK_{-}}  & e^{i K_{+}} - e^{iK_{-}}  \\
e^{i K_{+}} - e^{iK_{-}}  & e^{i K_{+}} + e^{iK_{-}} 
\end{array}
\right) ,
\label{eq.65}
\end{eqnarray}
which we have extended from the scalar form of the chain model~\cite{Hatano07,06TGP,05Sasada}.
At $x=\pm L \ (L>0) $, using the effective potential~(\ref{eq.65}), the left term of the time-dependent Schr\"odinger equation~(\ref{eq.61}) becomes 
\begin{eqnarray}
& \hat{\mathcal H}_{\textrm{eff}}&\vec{\Psi}_y (\pm L,t) \nonumber\\
 &=& -\frac{t_\textrm{h}}{2} \left\{ \vec{\Psi}_y (\pm (L-1),t) + \vec{\Psi}_y (\pm(L+1),t) \right\} \nonumber \\
& &  -t^{\prime}_\textrm{h} 
\left(
\begin{array}{cc}
0&1\\
1&0
\end{array}
\right) 
\vec{\Psi}_y (\pm L,t) \nonumber \\
&=& - \frac{t_\textrm{h}}{2} \vec{\Psi}_y (\pm(L-1),t)
      - \frac{t_\textrm{h}}{2} V_{\textrm{eff}} \vec{\Psi}_y (\pm L,t), \nonumber \\
& & - t^{\prime}_\textrm{h}S \vec{\Psi}_y (\pm L,t) , 
\label{eq.70} 
\end{eqnarray}
in which we define the matrix $S$ by
\begin{equation}
S\equiv 
\left(
\begin{array}{cc}
0&1\\
1&0
\end{array}
\right).
\label{eq.S}
\end{equation}

Thus, we can obtain the time-dependent Schr\"odinger equation in the \textit{closed region} $|x| \le L$ as follows:
\begin{eqnarray}
&&i \hbar \frac{\partial }{\partial t} \vec{\Psi}_y ( x,t) \nonumber \\
&&= 
\begin{cases}
	- \frac{t_\textrm{h}}{2} \left\{ \vec{\Psi}_y (\pm(L-1),t)
    +  V_{\textrm{eff}} \vec{\Psi}_y (\pm L,t) \right\} \\
    - t^{\prime}_\textrm{h}S	\vec{\Psi}_y (\pm L,t)   \\
	\qquad \qquad  	\qquad \qquad \qquad  \qquad \qquad \textrm{for \ \ } x=\pm L, \\
\\
	-\frac{t_\textrm{h}}{2} \left\{ \vec{\Psi}_y (x-1,t) + \vec{\Psi}_y (x+1,t) \right\}   \\
	\qquad -t^{\prime}_\textrm{h} S \vec{\Psi}_y (x,t) \\
	\qquad \qquad \qquad \qquad \qquad \qquad \textrm{for}\ 1 \le |x| \le L-1, \\
\\
	-\frac{t_\textrm{h}}{2} \left\{ \vec{\Psi}_y (-1,t) + \vec{\Psi}_y (1,t) \right\}  \\
	\qquad - t^{\prime}_\textrm{h} S \ \vec{\Psi}_y (0,t) 
	+g \Psi_{\textrm{d}} (t) 
	\begin{pmatrix}
	1 \\
	0
	\end{pmatrix} \\
	\qquad \qquad \qquad \qquad \qquad \qquad \qquad \qquad \textrm{for} \ x=0, \\
\end{cases}
\nonumber \\
& & \label{eq.80}
\end{eqnarray}
and 
\begin{equation}
i \hbar \frac{\partial }{\partial t} \Psi_{\textrm{d}} (t) =
E_{\textrm{d}}  \Psi_{\textrm{d}} (t) + g \Psi (0,1;t).\label{eq.90} 
\end{equation}
This is the method by which we have produced the time evolution simulations presented in Fig.~\ref{FIG-timeComp}.


\begin{thebibliography}{99}
\bibitem{footnote1}
Note that Eq.~(\ref{decoupled.ham}) represents the same Hamiltonian (shifted by the energy $t^\prime_\textrm{h}$) presented in the earlier publication~\cite{06TGP} on the single channel quantum wire.
\bibitem{footnote2}
Note that if we chose to obtain an expansion for the energy of the QBIC state R2 (instead of Q2) we would focus our attention on the embedded band edge at $t_\textrm{h} - t^\prime_\textrm{h}$ and in our discussion of the cause of the QBIC effect the band edges would play the opposite role, with $E_+$ giving the embedded band edge and $E_-$ being the embedding band.
\end{thebibliography}
\end{document}